\documentclass[%
reprint,
 amsmath,amssymb,
aps,
floatfix,
superscriptaddress,
showpacs,
prl
]{revtex4-2}

\usepackage{graphicx}
\usepackage{dcolumn}
\usepackage{bm}
\usepackage{color}
\usepackage[normalem]{ulem}
\usepackage{gensymb}

\graphicspath{{./}{figures/}}

\begin{document}

\preprint{APS/123-QED}

\title{Charge-Sign Dependent Cosmic-Ray Modulation Observed with the Calorimetric Electron Telescope on the International Space Station}

%
\author{O.~Adriani}
\affiliation{Department of Physics, University of Florence, Via Sansone, 1 - 50019, Sesto Fiorentino, Italy}
\affiliation{INFN Sezione di Firenze, Via Sansone, 1 - 50019, Sesto Fiorentino, Italy}
\author{Y.~Akaike}
\email[Corresponding author: ]{yakaike@aoni.waseda.jp}
\affiliation{Waseda Research Institute for Science and Engineering, Waseda University, 17 Kikuicho,  Shinjuku, Tokyo 162-0044, Japan}
\affiliation{JEM Utilization Center, Human Spaceflight Technology Directorate, Japan Aerospace Exploration Agency, 2-1-1 Sengen, Tsukuba, Ibaraki 305-8505, Japan}
\author{K.~Asano}
\affiliation{Institute for Cosmic Ray Research, The University of Tokyo, 5-1-5 Kashiwa-no-Ha, Kashiwa, Chiba 277-8582, Japan}
\author{Y.~Asaoka}
\affiliation{Institute for Cosmic Ray Research, The University of Tokyo, 5-1-5 Kashiwa-no-Ha, Kashiwa, Chiba 277-8582, Japan}
\author{E.~Berti} 
\affiliation{INFN Sezione di Firenze, Via Sansone, 1 - 50019, Sesto Fiorentino, Italy}
\affiliation{Institute of Applied Physics (IFAC),  National Research Council (CNR), Via Madonna del Piano, 10, 50019, Sesto Fiorentino, Italy}
\author{G.~Bigongiari}
\affiliation{Department of Physical Sciences, Earth and Environment, University of Siena, via Roma 56, 53100 Siena, Italy}
\affiliation{INFN Sezione di Pisa, Polo Fibonacci, Largo B. Pontecorvo, 3 - 56127 Pisa, Italy}
\author{W.R.~Binns}
\affiliation{Department of Physics and McDonnell Center for the Space Sciences, Washington University, One Brookings Drive, St. Louis, Missouri 63130-4899, USA}
\author{M.~Bongi}
\affiliation{Department of Physics, University of Florence, Via Sansone, 1 - 50019, Sesto Fiorentino, Italy}
\affiliation{INFN Sezione di Firenze, Via Sansone, 1 - 50019, Sesto Fiorentino, Italy}
\author{P.~Brogi}
\affiliation{Department of Physical Sciences, Earth and Environment, University of Siena, via Roma 56, 53100 Siena, Italy}
\affiliation{INFN Sezione di Pisa, Polo Fibonacci, Largo B. Pontecorvo, 3 - 56127 Pisa, Italy}
\author{A.~Bruno}
\affiliation{Heliospheric Physics Laboratory, NASA/GSFC, Greenbelt, Maryland 20771, USA}
\author{J.H.~Buckley}
\affiliation{Department of Physics and McDonnell Center for the Space Sciences, Washington University, One Brookings Drive, St. Louis, Missouri 63130-4899, USA}
\author{N.~Cannady}
\affiliation{Center for Space Sciences and Technology, University of Maryland, Baltimore County, 1000 Hilltop Circle, Baltimore, Maryland 21250, USA}
\affiliation{Astroparticle Physics Laboratory, NASA/GSFC, Greenbelt, Maryland 20771, USA}
\affiliation{Center for Research and Exploration in Space Sciences and Technology, NASA/GSFC, Greenbelt, Maryland 20771, USA}
\author{G.~Castellini}
\affiliation{Institute of Applied Physics (IFAC),  National Research Council (CNR), Via Madonna del Piano, 10, 50019, Sesto Fiorentino, Italy}
\author{C.~Checchia}
\affiliation{Department of Physical Sciences, Earth and Environment, University of Siena, via Roma 56, 53100 Siena, Italy}
\affiliation{INFN Sezione di Pisa, Polo Fibonacci, Largo B. Pontecorvo, 3 - 56127 Pisa, Italy}
\author{M.L.~Cherry}
\affiliation{Department of Physics and Astronomy, Louisiana State University, 202 Nicholson Hall, Baton Rouge, Louisiana 70803, USA}
\author{G.~Collazuol}
\affiliation{Department of Physics and Astronomy, University of Padova, Via Marzolo, 8, 35131 Padova, Italy}
\affiliation{INFN Sezione di Padova, Via Marzolo, 8, 35131 Padova, Italy} 
\author{G.A.~de~Nolfo}
\affiliation{Heliospheric Physics Laboratory, NASA/GSFC, Greenbelt, Maryland 20771, USA}
\author{K.~Ebisawa}
\affiliation{Institute of Space and Astronautical Science, Japan Aerospace Exploration Agency, 3-1-1 Yoshinodai, Chuo, Sagamihara, Kanagawa 252-5210, Japan}
\author{A.~W.~Ficklin}
\affiliation{Department of Physics and Astronomy, Louisiana State University, 202 Nicholson Hall, Baton Rouge, Louisiana 70803, USA}
\author{H.~Fuke}
\affiliation{Institute of Space and Astronautical Science, Japan Aerospace Exploration Agency, 3-1-1 Yoshinodai, Chuo, Sagamihara, Kanagawa 252-5210, Japan}
\author{S.~Gonzi}
\affiliation{Department of Physics, University of Florence, Via Sansone, 1 - 50019, Sesto Fiorentino, Italy}
\affiliation{INFN Sezione di Firenze, Via Sansone, 1 - 50019, Sesto Fiorentino, Italy}
\affiliation{Institute of Applied Physics (IFAC),  National Research Council (CNR), Via Madonna del Piano, 10, 50019, Sesto Fiorentino, Italy}
\author{T.G.~Guzik}
\affiliation{Department of Physics and Astronomy, Louisiana State University, 202 Nicholson Hall, Baton Rouge, Louisiana 70803, USA}
\author{T.~Hams}
\affiliation{Center for Space Sciences and Technology, University of Maryland, Baltimore County, 1000 Hilltop Circle, Baltimore, Maryland 21250, USA}
\author{K.~Hibino}
\affiliation{Kanagawa University, 3-27-1 Rokkakubashi, Kanagawa, Yokohama, Kanagawa 221-8686, Japan}
\author{M.~Ichimura}
\affiliation{Faculty of Science and Technology, Graduate School of Science and Technology, Hirosaki University, 3, Bunkyo, Hirosaki, Aomori 036-8561, Japan}
\author{K.~Ioka}
\affiliation{Yukawa Institute for Theoretical Physics, Kyoto University, Kitashirakawa Oiwake-cho, Sakyo-ku, Kyoto, 606-8502, Japan}
\author{W.~Ishizaki}
\affiliation{Institute for Cosmic Ray Research, The University of Tokyo, 5-1-5 Kashiwa-no-Ha, Kashiwa, Chiba 277-8582, Japan}
\author{M.H.~Israel}
\affiliation{Department of Physics and McDonnell Center for the Space Sciences, Washington University, One Brookings Drive, St. Louis, Missouri 63130-4899, USA}
\author{K.~Kasahara}
\affiliation{Department of Electronic Information Systems, Shibaura Institute of Technology, 307 Fukasaku, Minuma, Saitama 337-8570, Japan}
\author{J.~Kataoka}
\affiliation{School of Advanced Science and	Engineering, Waseda University, 3-4-1 Okubo, Shinjuku, Tokyo 169-8555, Japan}
\author{R.~Kataoka}
\affiliation{National Institute of Polar Research, 10-3, Midori-cho, Tachikawa, Tokyo 190-8518, Japan}
\author{Y.~Katayose}
\affiliation{Faculty of Engineering, Division of Intelligent Systems Engineering, Yokohama National University, 79-5 Tokiwadai, Hodogaya, Yokohama 240-8501, Japan}
\author{C.~Kato}
\affiliation{Faculty of Science, Shinshu University, 3-1-1 Asahi, Matsumoto, Nagano 390-8621, Japan}
\author{N.~Kawanaka}
\affiliation{Yukawa Institute for Theoretical Physics, Kyoto University, Kitashirakawa Oiwake-cho, Sakyo-ku, Kyoto, 606-8502, Japan}
\author{Y.~Kawakubo}
\affiliation{Department of Physics and Astronomy, Louisiana State University, 202 Nicholson Hall, Baton Rouge, Louisiana 70803, USA}
\author{K.~Kobayashi}
\affiliation{Waseda Research Institute for Science and Engineering, Waseda University, 17 Kikuicho,  Shinjuku, Tokyo 162-0044, Japan}
\affiliation{JEM Utilization Center, Human Spaceflight Technology Directorate, Japan Aerospace Exploration Agency, 2-1-1 Sengen, Tsukuba, Ibaraki 305-8505, Japan}
\author{K.~Kohri} 
\affiliation{Institute of Particle and Nuclear Studies, High Energy Accelerator Research Organization, 1-1 Oho, Tsukuba, Ibaraki, 305-0801, Japan} 
\author{H.S.~Krawczynski}
\affiliation{Department of Physics and McDonnell Center for the Space Sciences, Washington University, One Brookings Drive, St. Louis, Missouri 63130-4899, USA}
\author{J.F.~Krizmanic}
\affiliation{Astroparticle Physics Laboratory, NASA/GSFC, Greenbelt, Maryland 20771, USA}
\author{P.~Maestro}
\affiliation{Department of Physical Sciences, Earth and Environment, University of Siena, via Roma 56, 53100 Siena, Italy}
\affiliation{INFN Sezione di Pisa, Polo Fibonacci, Largo B. Pontecorvo, 3 - 56127 Pisa, Italy}
\author{P.S.~Marrocchesi}
\affiliation{Department of Physical Sciences, Earth and Environment, University of Siena, via Roma 56, 53100 Siena, Italy}
\affiliation{INFN Sezione di Pisa, Polo Fibonacci, Largo B. Pontecorvo, 3 - 56127 Pisa, Italy}
\author{A.M.~Messineo}
\affiliation{INFN Sezione di Pisa, Polo Fibonacci, Largo B. Pontecorvo, 3 - 56127 Pisa, Italy}
\affiliation{University of Pisa, Polo Fibonacci, Largo B. Pontecorvo, 3 - 56127 Pisa, Italy}
\author{J.W.~Mitchell}
\affiliation{Astroparticle Physics Laboratory, NASA/GSFC, Greenbelt, Maryland 20771, USA}
\author{S.~Miyake}
\email[Corresponding author: ]{miyakesk@ee.ibaraki-ct.ac.jp}
\affiliation{Department of Electrical and Electronic Systems Engineering, National Institute of Technology (KOSEN), Ibaraki College, 866 Nakane, Hitachinaka, Ibaraki 312-8508, Japan}
\author{A.A.~Moiseev}
\affiliation{Astroparticle Physics Laboratory, NASA/GSFC, Greenbelt, Maryland 20771, USA}
\affiliation{Center for Research and Exploration in Space Sciences and Technology, NASA/GSFC, Greenbelt, Maryland 20771, USA}
\affiliation{Department of Astronomy, University of Maryland, College Park, Maryland 20742, USA}\author{M.~Mori}
\affiliation{Department of Physical Sciences, College of Science and Engineering, Ritsumeikan University, Shiga 525-8577, Japan}
\author{N.~Mori}
\affiliation{INFN Sezione di Firenze, Via Sansone, 1 - 50019, Sesto Fiorentino, Italy}
\author{H.M.~Motz}
\affiliation{Faculty of Science and Engineering, Global Center for Science and Engineering, Waseda University, 3-4-1 Okubo, Shinjuku, Tokyo 169-8555, Japan}
\author{K.~Munakata}
\email[Corresponding author: ]{kmuna00@shinshu-u.ac.jp}
\affiliation{Faculty of Science, Shinshu University, 3-1-1 Asahi, Matsumoto, Nagano 390-8621, Japan}
\author{S.~Nakahira}
\affiliation{Institute of Space and Astronautical Science, Japan Aerospace Exploration Agency, 3-1-1 Yoshinodai, Chuo, Sagamihara, Kanagawa 252-5210, Japan}
\author{J.~Nishimura}
\affiliation{Institute of Space and Astronautical Science, Japan Aerospace Exploration Agency, 3-1-1 Yoshinodai, Chuo, Sagamihara, Kanagawa 252-5210, Japan}
\author{S.~Okuno}
\affiliation{Kanagawa University, 3-27-1 Rokkakubashi, Kanagawa, Yokohama, Kanagawa 221-8686, Japan}
\author{J.F.~Ormes}
\affiliation{Department of Physics and Astronomy, University of Denver, Physics Building, Room 211, 2112 East Wesley Avenue, Denver, Colorado 80208-6900, USA}
\author{S.~Ozawa}
\affiliation{Quantum ICT Advanced Development Center, National Institute of Information and Communications Technology, 4-2-1 Nukui-Kitamachi, Koganei, Tokyo 184-8795, Japan}
\author{L.~Pacini}
\affiliation{INFN Sezione di Firenze, Via Sansone, 1 - 50019, Sesto Fiorentino, Italy}
\affiliation{Institute of Applied Physics (IFAC),  National Research Council (CNR), Via Madonna del Piano, 10, 50019, Sesto Fiorentino, Italy}
\author{P.~Papini}
\affiliation{INFN Sezione di Firenze, Via Sansone, 1 - 50019, Sesto Fiorentino, Italy}
\author{B.F.~Rauch}
\affiliation{Department of Physics and McDonnell Center for the Space Sciences, Washington University, One Brookings Drive, St. Louis, Missouri 63130-4899, USA}
\author{S.B.~Ricciarini}
\affiliation{INFN Sezione di Firenze, Via Sansone, 1 - 50019, Sesto Fiorentino, Italy}
\affiliation{Institute of Applied Physics (IFAC),  National Research Council (CNR), Via Madonna del Piano, 10, 50019, Sesto Fiorentino, Italy}\author{K.~Sakai}
\affiliation{Center for Space Sciences and Technology, University of Maryland, Baltimore County, 1000 Hilltop Circle, Baltimore, Maryland 21250, USA}
\affiliation{Astroparticle Physics Laboratory, NASA/GSFC, Greenbelt, Maryland 20771, USA}
\affiliation{Center for Research and Exploration in Space Sciences and Technology, NASA/GSFC, Greenbelt, Maryland 20771, USA}
\author{T.~Sakamoto}
\affiliation{College of Science and Engineering, Department of Physics and Mathematics, Aoyama Gakuin University,  5-10-1 Fuchinobe, Chuo, Sagamihara, Kanagawa 252-5258, Japan}
\author{M.~Sasaki}
\affiliation{Astroparticle Physics Laboratory, NASA/GSFC, Greenbelt, Maryland 20771, USA}
\affiliation{Center for Research and Exploration in Space Sciences and Technology, NASA/GSFC, Greenbelt, Maryland 20771, USA}
\affiliation{Department of Astronomy, University of Maryland, College Park, Maryland 20742, USA}\author{Y.~Shimizu}
\affiliation{Kanagawa University, 3-27-1 Rokkakubashi, Kanagawa, Yokohama, Kanagawa 221-8686, Japan}
\author{A.~Shiomi}
\affiliation{College of Industrial Technology, Nihon University, 1-2-1 Izumi, Narashino, Chiba 275-8575, Japan}
\author{P.~Spillantini}
\affiliation{Department of Physics, University of Florence, Via Sansone, 1 - 50019, Sesto Fiorentino, Italy}
\author{F.~Stolzi}
\affiliation{Department of Physical Sciences, Earth and Environment, University of Siena, via Roma 56, 53100 Siena, Italy}
\affiliation{INFN Sezione di Pisa, Polo Fibonacci, Largo B. Pontecorvo, 3 - 56127 Pisa, Italy}
\author{S.~Sugita}
\affiliation{College of Science and Engineering, Department of Physics and Mathematics, Aoyama Gakuin University,  5-10-1 Fuchinobe, Chuo, Sagamihara, Kanagawa 252-5258, Japan}
\author{A.~Sulaj} 
\affiliation{Department of Physical Sciences, Earth and Environment, University of Siena, via Roma 56, 53100 Siena, Italy}
\affiliation{INFN Sezione di Pisa, Polo Fibonacci, Largo B. Pontecorvo, 3 - 56127 Pisa, Italy}
\author{M.~Takita}
\affiliation{Institute for Cosmic Ray Research, The University of Tokyo, 5-1-5 Kashiwa-no-Ha, Kashiwa, Chiba 277-8582, Japan}
\author{T.~Tamura}
\affiliation{Kanagawa University, 3-27-1 Rokkakubashi, Kanagawa, Yokohama, Kanagawa 221-8686, Japan}
\author{T.~Terasawa}
\affiliation{Institute for Cosmic Ray Research, The University of Tokyo, 5-1-5 Kashiwa-no-Ha, Kashiwa, Chiba 277-8582, Japan}
\author{S.~Torii}
\affiliation{Waseda Research Institute for Science and Engineering, Waseda University, 17 Kikuicho,  Shinjuku, Tokyo 162-0044, Japan}
\author{Y.~Tsunesada}
\affiliation{Graduate School of Science, Osaka Metropolitan University, Sugimoto, Sumiyoshi, Osaka 558-8585, Japan }
\affiliation{Nambu Yoichiro Institute for Theoretical and Experimental Physics, Osaka Metropolitan University,  Sugimoto, Sumiyoshi, Osaka  558-8585, Japan}
\author{Y.~Uchihori}
\affiliation{National Institutes for Quantum and Radiation Science and Technology, 4-9-1 Anagawa, Inage, Chiba 263-8555, Japan}
\author{E.~Vannuccini}
\affiliation{INFN Sezione di Firenze, Via Sansone, 1 - 50019, Sesto Fiorentino, Italy}
\author{J.P.~Wefel}
\affiliation{Department of Physics and Astronomy, Louisiana State University, 202 Nicholson Hall, Baton Rouge, Louisiana 70803, USA}
\author{K.~Yamaoka}
\affiliation{Nagoya University, Furo, Chikusa, Nagoya 464-8601, Japan}
\author{S.~Yanagita}
\affiliation{College of Science, Ibaraki University, 2-1-1 Bunkyo, Mito, Ibaraki 310-8512, Japan}
\author{A.~Yoshida}
\affiliation{College of Science and Engineering, Department of Physics and Mathematics, Aoyama Gakuin University,  5-10-1 Fuchinobe, Chuo, Sagamihara, Kanagawa 252-5258, Japan}
\author{K.~Yoshida}
\affiliation{Department of Electronic Information Systems, Shibaura Institute of Technology, 307 Fukasaku, Minuma, Saitama 337-8570, Japan}
\author{W.~V.~Zober}
\affiliation{Department of Physics and McDonnell Center for the Space Sciences, Washington University, One Brookings Drive, St. Louis, Missouri 63130-4899, USA}

\collaboration{CALET Collaboration}


\begin{abstract}
We present the observation of a charge-sign dependent solar modulation of galactic cosmic rays (GCRs) with the CALorimetric Electron Telescope onboard the International Space Station over 6 yr, corresponding to the positive polarity of the solar magnetic field. 
The observed variation of proton count rate is consistent with the neutron monitor count rate, validating our methods for determining the proton count rate. 
It is observed by the CALorimetric Electron Telescope that both GCR electron and proton count rates at the same average rigidity vary in anticorrelation with the tilt angle of the heliospheric current sheet, while the amplitude of the variation is significantly larger in the electron count rate than in the proton count rate. 
We show that this observed charge-sign dependence is reproduced by a numerical ``drift model'' of the GCR transport in the heliosphere. This is a clear signature of the drift effect on the long-term solar modulation observed with a single detector.

\end{abstract}

\maketitle


\section{\label{sec:intro}Introduction}

The galactic cosmic ray (GCR) intensity observed at Earth shows a clear $\sim$11-yr cycle variation in anticorrelation with the solar activity. 
This well-known phenomenon known as the heliospheric modulation of GCRs has been interpreted as a result of the large-scale GCR transport in the heliosphere. 
The potential importance of the gradient and curvature drift in the GCR transport was addressed theoretically by \citet{JLH1977}. 
Numerical calculations by \citet{JT1981} and \citet{KJ1983} showed that the drift effect results in an anticorrelation between the GCR intensity at Earth and the tilt angle of the heliospheric current sheet (HCS) which increases with the solar activity and the waviness of the HCS. 
Since the orientation of the guiding center drift velocity reverses according to the alteration of the sign ($q$) of the particle's charge and the sign ($A$) of the solar magnetic field polarity, the drift effect is predicted to have different anticorrelations between the GCR intensity and the HCS tilt angle when $qA>0$ or $qA<0$. 
During periods with $A>0$ the solar polar magnetic field is directed away from (toward) the Sun in the northern (southern) hemisphere, and the drift leads electrons ($q<0$) inward toward the Earth along the HCS while the distance along the access route becomes longer as HCS waviness increases. 
The drift, on the other hand, leads protons ($q>0$) to arrive at Earth from the heliospheric polar region, reaching the HCS less often. 
This results in a larger modulation of the electron flux than that of the proton flux at Earth given the HCS tilt angle increase during periods with $A>0$. 
The same effect is also seen in ``peaked'' and ``flat'' maxima of $\sim$11-yr GCR variations when $qA<0$ or $qA>0$, respectively \citep{KJ1983}.\\

So far, the drift effect has been explored by  analyzing GCR data in several ways. 
By examining the anticorrelation between the HCS tilt angle and the neutron monitor count rates corresponding to GCR protons and helium with $q>0$, \citet{Cane1999} found no clear difference in the anticorrelations when $A>0$ or $A<0$ except for short periods around the solar activity minima. 
They attributed the majority of the observed modulation to the change of the Sun's magnetic field strength instead of the change of the HCS tilt angle. 
In such an analysis with data taken at different time periods, however, it is difficult to distinguish the difference due to the drift effect from other modulation parameters.
For example, other parameters include the solar wind velocity and the magnetic field strength, which are generally not the same in different time periods \citep{ACE04}. 
\citet{Bieber1999}, on the other hand, suggested that a notable change in the flux ratio of GCRs with $q< 0$ or $q> 0$ is expected from the drift model and would provide a good test of the magnitude of the effect. 
This change is actually observed in the flux ratio ($\bar{p}/p$) of GCR antiprotons ($\bar{p}$) and protons ($p$) from the BESS balloon-borne experiment \citep{Asaoka2002}. 
The observed change of $\bar{p}/p$ is also quantitatively reproduced by the numerical calculation of the drift model. 
Recently, \citet{Pamela2016} and \citet{AMSe2018} also reported similar results from analyzing the flux ratio ($e^{+}/e^{-}$) of GCR positrons ($e^{+}$) and electrons ($e^{-}$) observed by several space experiments. 
However, the $qA$ dependence of the anticorrelation between the GCR intensity and the HCS tilt angle over the solar activity cycle has not been reported yet. 
In this Letter, we report for the first time the anticorrelations with the HCS tilt angle of the electron and proton count rates simultaneously observed by the CALorimetric Electron Telescope (CALET) \citep{CALope2017, CALe2017, CALET2017torii, CALope2018, CALe2018, CALp2019, CALET2019torii, CALET2019asaoka, CALco2020, CALp2022} onboard the International Space Station over nearly 6 yr between 2015 and 2021.\\

\section{\label{sec:obsana}CALET instrument and data analysis}

\begin{figure}[t]
\includegraphics[width=17pc]{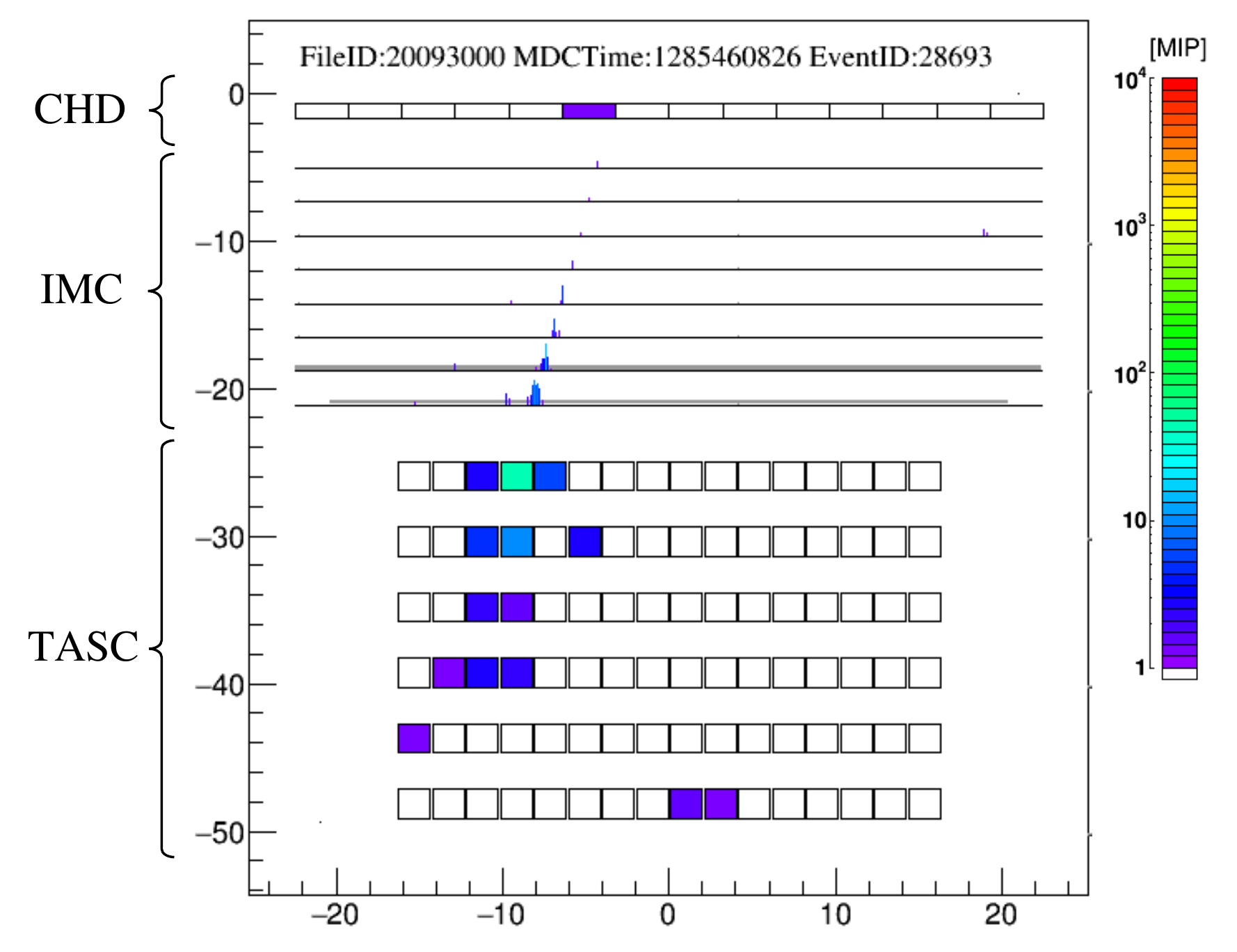}
\caption{\label{fig_instrument}Schematic cross-sectional view of CALET with an electron LE event candidate with energy of $\sim 3.9~\mathrm{GeV}$.}
\end{figure}

Figure 1 shows a schematic side view of CALET consisting of a charge detector (CHD) for identifying the charge of the incident particle, an imaging calorimeter (IMC) for track reconstruction and for fine-spatial resolution imaging of the early stage shower development, and a total absorption calorimeter (TASC) for measuring the energy of the electromagnetic shower \citep{CALope2018}. 
The CHD is composed of a pair of $x$-$y$ layers each consisting of 14 plastic scintillator paddles with dimensions of 450 mm long~$\times$~32 mm wide~$\times$~10 mm thick. The IMC is composed of eight $x$-$y$ layers of 448 mm long~$\times$~1 mm square cross-section scintillating fibers interleaved with tungsten plates. 
The first five tungsten plates have 0.2 radiation length ($X_0$) thickness and the last two plates each have 1.0$X_0$ thickness. 
The total thickness of the IMC is equivalent to 3$X_0$. 
The TASC consists of 12 crossed layers of 16 lead tungstate logs, each with dimensions of 326 mm long~$\times$~19 mm wide~$\times$~20 mm thick, for a total thickness of 27$X_0$. 
The total thickness of the calorimeter is 30 $X_0$, equivalent to $\sim$1.3 proton interaction lengths, allowing CALET to obtain near total absorption of electron showers even up to about 20 TeV.

The normal event trigger of CALET is provided by the high-energy trigger with an energy threshold of $\sim$10 GeV. 
In addition to the high-energy trigger, in order to collect the low-energy ($\gtrsim 1~\mathrm{GeV}$) particle events efficiently, a low-energy electron (LEE) trigger is useful at high geographical latitudes where the geomagnetic cutoff rigidity (COR) is below 5.0 GV. 
This LEE shower trigger mode is operated for 90 s twice per International Space Station orbital period ($\sim91$ m), at a $51.6^\circ$ orbital inclination, in each of the north and south regions. In this study, we analyze the flight data collected in the LEE trigger mode during 2058 d from October 13, 2015 to May 31, 2021.
We have collected about $91\times10^6$ low-energy GCR candidates in a total observational live time of approximately 766 h. 
From this dataset, we select electrons and protons and deduce their count rates for the same average rigidity.  

For the event selection and energy reconstruction, we adopt a Monte Carlo (MC) simulation developed to simulate physical processes and detector response based on the simulation package EPICS \citep{EPICS} (EPICS 9.20 and COSMOS 8.00) and the DPMJET-III model for hadron interactions. 
The MC event samples consist of the response to downward propagating electron and proton events generated isotropically on a spherical surface with a radius of 78 cm surrounding the instrument. 
We apply the following event-selection criteria: (a) off-line trigger condition requiring energy deposits in the bottom two layers of the IMC and the top layer of the TASC to exceed a given set of thresholds, (b) quality cut on the  reconstructed track of the incident particle by the Kalman filter method, (c) geometrical condition requiring the reconstructed track to traverse the CHD top layer and the TASC bottom layer, (d) cut on the CHD output to select incident particles with single charge, (e) cut on an energy deposit in all layers of the IMC and the TASC to exclude events passing through the layer without energy deposit, (f) additional cut on the spatial concentration of hit signals in the IMC bottom layer to reduce the proton contamination for the analysis of electron count rates, and (g) cut on the lateral shower development in the TASC top layer for electron and proton discrimination [see the Supplemental Material \citep{SM} about the detail of criteria (f) and (g)]. 
Details of these criteria are provided in \citep{CALe2017, CALe2018} for the analysis of high-energy electrons, with the important distinction that the analysis here imposes selections on the IMC bottom layer and TASC top layer for electron and proton discrimination given that the low-energy electrons do not penetrate all layers of the TASC. 

In order to minimize the count rate variation due to the COR, we choose periods in which the COR is below 0.8 GV and select events recorded with a deposited energy exceeding 1.0 GeV.
The COR is calculated by back-tracing the particle trajectory in the magnetosphere defined by the IGRF-13 \citep{IGRF13} and TS05 \citep{Tsyg04} empirical models \citep{Miyake2017} for every incident direction.
Orbit calculations are repeated by decreasing the particle's rigidity and the COR is defined as the lowest rigidity before the appearance of the penumbra.

For the analysis of the charge-sign dependent solar modulation in this Letter, we derive the count rates of electrons and protons at the same average rigidity. 
The average rigidity of electrons that passed the above selection criteria is estimated to be $\sim3.8$ GV from MC simulations. The average rigidity of protons is adjusted to $\sim3.8$ GV by selecting the events for which the energy deposit in all layers of the IMC and the TASC is between 1 and 3 GeV, which is verified from MC simulations.
We analyze about $0.77\times10^6$ electron and $1.26\times10^6$ proton candidates collected in a total observational live time of about 196 and 197 h, respectively.

\section{Results and discussions \label{sec:result}}

\begin{figure}[t]
\centering
\includegraphics[width=17pc]{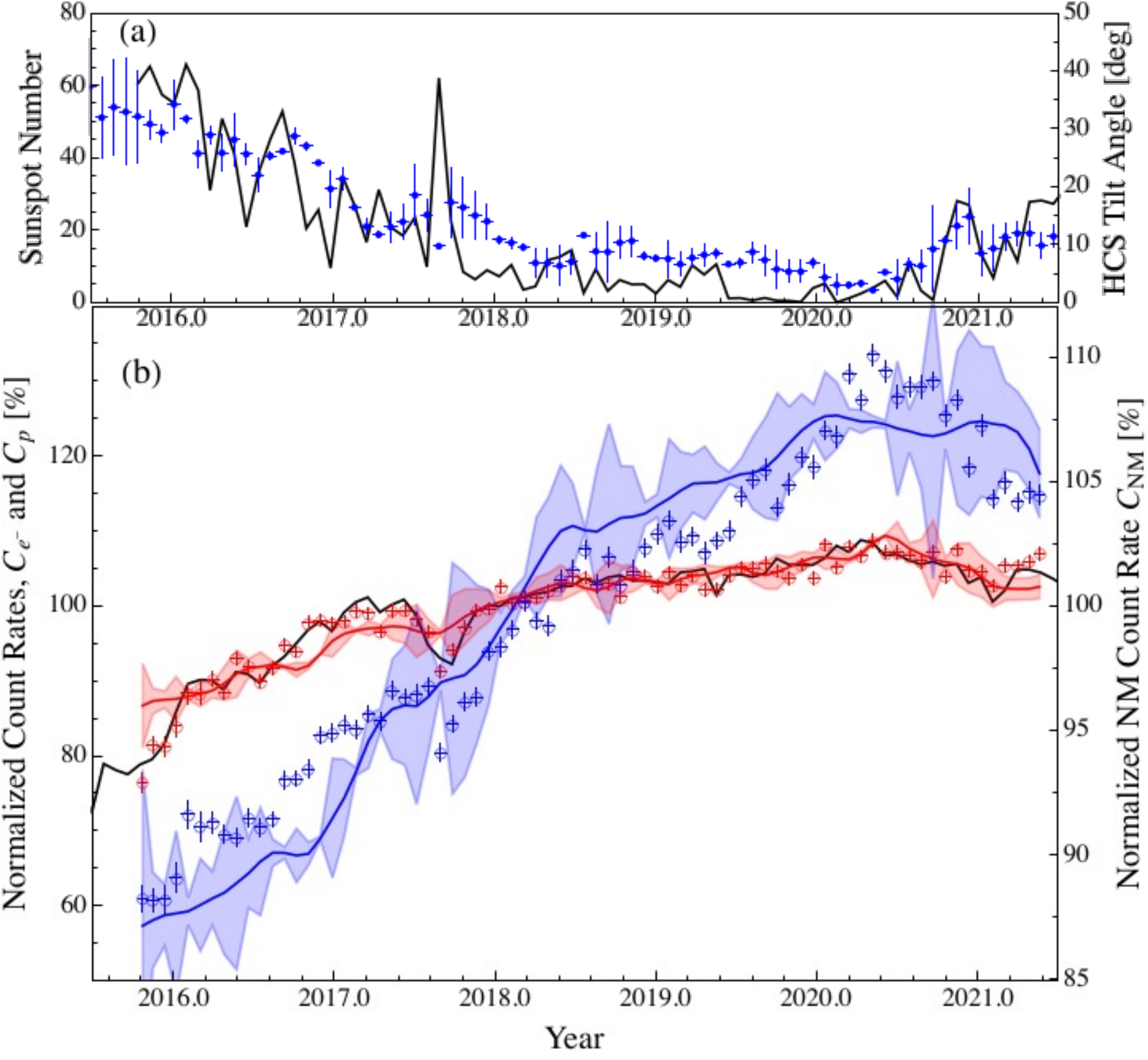}
\caption{\label{fig_charge-dep} (a) Time profiles of the sunspot number (solid line) and the HCS tilt angles (blue solid circles) as a function of the central time of each Carrington rotation. (b) Time profiles of the normalized count rates of electrons $C_{e^-}$ (blue open circles) and protons $C_p$ (red open circles) for each Carrington rotation (left vertical axis), where statistical uncertainties are shown. 
The black curve shows the count rate of a neutron monitor at the Oulu station on the right vertical axis, while the blue and red curves show the electron and proton count rates reproduced by the numerical model, respectively. Each shaded area around the reproduced curve indicates the error deduced from the error of the HCS tilt angle and the regression coefficient between the tilt angle and the reproduced curve (see text).}
\end{figure}

Figure$~$2(b) shows the electron and proton count rates at an average rigidity of 3.8 GV (blue and red symbols), $C_{e^-}$ and $C_{p}$, respectively, observed by CALET for the 6-yr period corresponding to $A>0$ \citep{Pol_WSOweb}. 
Shown in the figure are only statistical uncertainties given that the systematic uncertainties do not vary appreciably over the period under consideration.  
For reference, Fig.$~$2(a) displays the sunspot number (SSN) \citep{OMNIweb} and the HCS tilt angle based on the radial model at the Wilcox Solar Observatory \citep{HCS_WSOweb} (courtesy of \citet{HCS_WSO}) representing the solar activity in the same period. 
The error on the HCS tilt angle is deduced from the difference between the values in the northern and southern hemispheres.
In Fig.$~$2(b), the average count rate over the entire period is normalized to 100 and each symbol shows an average within each solar rotation (Carrington rotation) period. 
Since CALET is incapable of discriminating positrons from electrons, it is necessary to correct the temporal variation of the observed electron flux for the contribution from the positron flux which has been found to be below $10\%$ of the electron flux at $3.8$ GV \citep{Pamela2016, AMSe2018}.
We correct $C_{e^-}$ for this positron contamination by assuming that the positron contamination is $7\%$ on October 2015 \citep{AMSe2018} and that the variation of the normalized positron count rate is identical to $C_{p}$ in Fig.$~$2(b) (see the Supplemental Material \citep{SM} for more detail). 
We find that the amplitude of the $C_{e^-}$ variation increases by about $3.1\%$ by taking this correction into account. 

The electron count rate measured by CALET reached its maximum about 6 months after the beginning of solar cycle 25 in December 2019 [see SSN and HCS tilt angle in Fig.$~$2(a)]. 
Also shown in Fig.$~$2(b) is the Oulu neutron monitor count rate ($C_{\mathrm{NM}}$) \citep{Oule,OuluNM} which is sensitive to high-energy ($\sim$ 10 GV) GCR protons (black solid curve). 
A good correlation is seen between $C_{p}$ and $C_{\mathrm{NM}}$ with a correlation coefficient of 0.98 [see Fig.$~$S2 of the Supplemental Material \citep{SM}]. 
The ratio of $C_{p}$ to $C_{\mathrm{NM}}$ with average rigidity of $\sim$10 GV is about 3.32, indicating the rigidity spectral index of the proton count rates is about $-1.24$.
This result is consistent with the spectrum known for the long-term solar modulation in this rigidity region (e.g., Munakata {\it et al.} \cite{MK2019}), providing further support that our determination of the proton count rate is handled correctly. 

\begin{figure}[t]
\centering
\includegraphics[width=17pc]{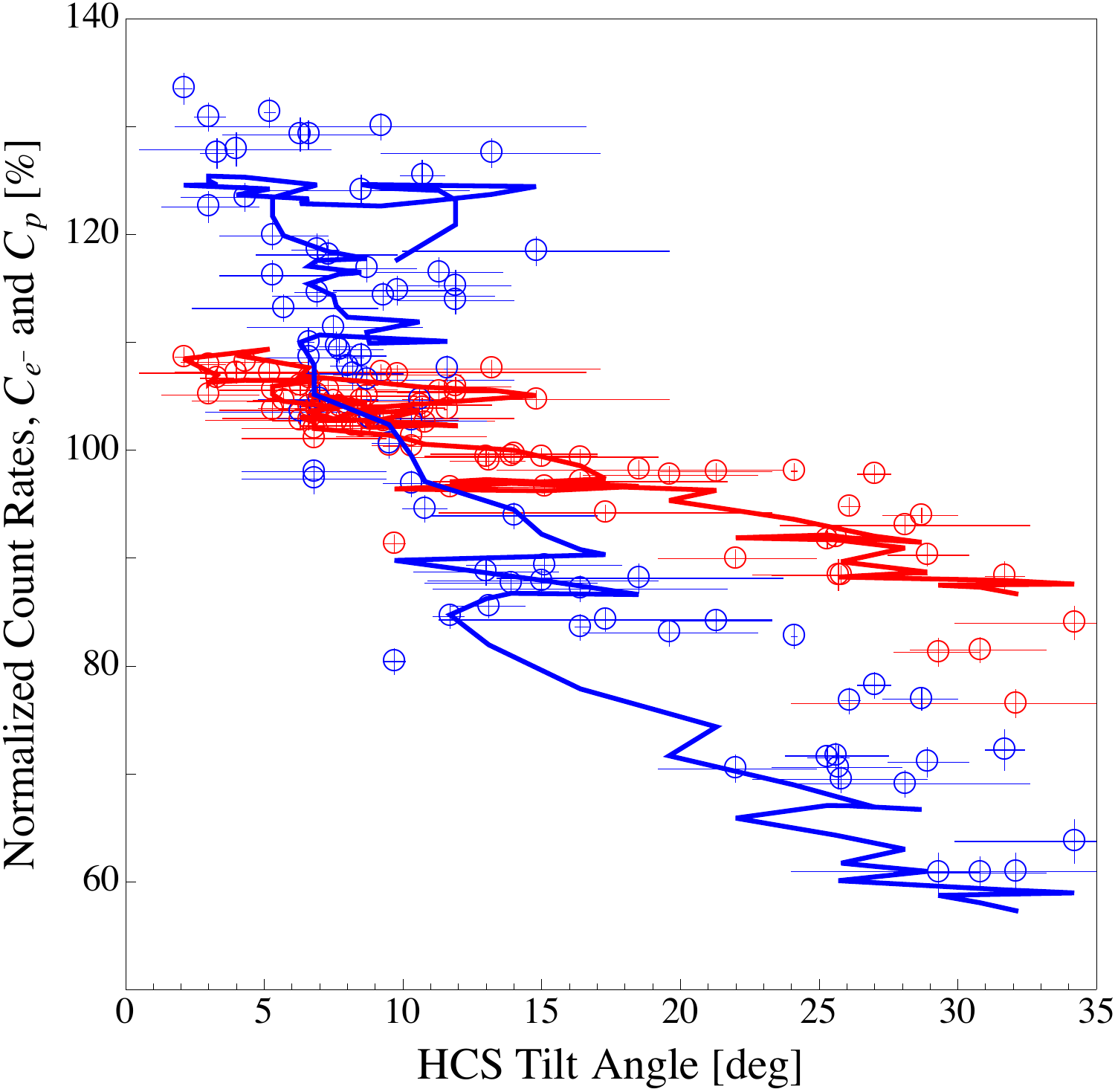}
\caption{\label{fig_vsHCS}Correlation plot between the normalized count rate and the HCS tilt angle. Blue (red) open circles show the correlation of the observed electron (proton) count rate, respectively, while blue and red solid curves display correlations reproduced by the numerical model (see text).}
\end{figure}

From Fig.$~$2(b), we see that both $C_{e^-}$ and $C_{p}$ increase with decreasing solar activity toward the solar minimum in 2019, as indicated by the SSN and HCS tilt angle in Fig.$~$2(a). 
The most striking feature in Fig.$~$2(b) is that the variation of $C_{e^-}$ is clearly larger than that of $C_{p}$ at the same average rigidity. 
As mentioned in the introduction, this is consistent with the drift effect in which a stronger anticorrelation between the GCR intensity and the HCS tilt angle results for $qA < 0$ than for $qA > 0$. 
The blue (red) solid curve displays $C_{e^-}$ ($C_{p}$) calculated by a numerical drift model \citep{Miyake2017,MiyakeICRC2017} (see the Supplemental Material \citep{SM} for details), with an average uncertainty of $5.9 \pm 4.3$ ($\%$) ($1.6 \pm 1.2$ ($\%$)) mainly due to the error in the HCS tilt angle. 
The model simultaneously reproduces the observed variations in both $C_{e^-}$ and $C_{p}$, with some departures in the predictions of $C_{e^-}$, most notably before 2018.
 
Figure 3 shows the observed $C_{e^-}$ (blue symbols) and $C_{p}$ (red symbols) as a function of the HCS tilt angle on the horizontal axis together with the model prediction displayed by the blue (red) solid curve. 
The regression coefficient of the observed $C_{e^-}$ as a function of the HCS tilt angle is $-2.12 \pm 0.17$ ($\%/^\circ$) and is significantly larger than $-0.72 \pm 0.06$ ($\%/^\circ$) for $C_{p}$. 
The regression coefficients of the reproduced $C_{e^-}$ and $C_{p}$ are $-2.57 \pm 0.19$ ($\%/^\circ$) and $-0.69 \pm 0.05$ ($\%/^\circ$), respectively, roughly consistent with our observations. 
Differences between the observed and modeled correlation with the HCS tilt angle can be attributed to model-dependent assumptions that can be further refined in the future.  
For instance, the model could better represent distortions in the HCS introduced by solar wind disturbances, including coronal interaction regions \citep{Richardson2018}.
It is also known that the GCR variation at Earth lags several Carrington rotations behind the HCS tilt angle, as seen in Fig.$~$2 by the fact that the maxima of $C_{e^-}$ and $C_{p}$ are delayed with respect to the minimum value of the HCS tilt angle. 
This hysteresis effect is shown in Fig.$~$3 with clockwise rotations of $C_{e^-}$ and $C_{p}$ that are also reproduced by the numerical model (solid curves in Fig.$~$3). 

In summary, we have determined the solar rotation averages of electron ($q<0$) and proton ($q>0$) count rates, $C_{e^-}$ and $C_{p}$, measured by CALET at the same average rigidity for approximately 6 yr from 2015 to 2020 corresponding to $A>0$ of the solar magnetic field polarity. 
A good correlation between $C_{p}$ and $C_{\mathrm{NM}}$ with a correlation coefficient of 0.98 validates the determination of the CALET proton count rate. 
It is found that the modulation amplitude of the average $C_{e^-}$ is clearly larger than that of $C_{p}$, consistent with the drift model predictions of a larger anticorrelation between the GCR intensity and the HCS tilt angle when $qA<0$ than that with $qA>0$. 
It is also shown that the observed modulations of $C_{e^-}$ and $C_{p}$ are simultaneously reproduced by a numerical model that accounts for the drift effect in the GCR transport in the heliosphere. 
This is the first clear evidence of the drift effect playing a major role in the long-term modulation of GCRs.

\begin{acknowledgments}
We used JAXA Supercomputer System generation 3 (JSS3) with business name: ``Numerical study on the solar modulation of low-energy cosmic rays measured with CALET" (business code: ACA51). 
We gratefully acknowledge JAXA’s contributions to the development of CALET and to the operations onboard the International Space Station. We also express our sincere gratitude to ASI and NASA for their support of the CALET project. 
This work was supported in part by JSPS Grant-in-Aid for Scientific Research (S) Grant No. 19H05608, JSPS Grant-in-Aid for Scientific Research (C) Grant No. 20K03956, JSPS Grant-in-Aid for Scientific Research (C) Grant No. 21K03592, and by the MEXT-Supported Program for the Strategic Research Foundation at Private Universities (2011-2015) (Grant No. S1101021) at Waseda University. The CALET effort in Italy is supported by ASI under Agreement No. 2013-018-R.0 and its amendments. The CALET effort in the U.S. is supported by NASA through Grants No. 80NSSC20K0397, No. 80NSSC20K0399, No. NNH18ZDA001N-APRA18-004, and under Award No. 80GSFC21M0002.
\end{acknowledgments}




\nocite{*}

\bibliography{caletSMdraft_ver17}

\end{document}



\title{Charge-Sign Dependent Cosmic-Ray Modulation Observed with the Calorimetric Electron Telescope on the International Space Station\\
SUPPLEMENTAL MATERIAL}

%
\collaboration{CALET Collaboration}
\maketitle

Supplemental material relative to ``Charge-Sign Dependent Cosmic-Ray Modulation Observed with the Calorimetric Electron Telescope on the International Space Station."


\section{Background Contamination}
In each Carrington rotation period, we classify events that passed the event selection criteria (a)-(e) described in the main text into electrons and positrons (hereafter ${e}^{+/-}$) and protons (p) with two additional criteria (f) and (g). 
Criterion (f) consists of a selection performed only for ${e}^{+/-}$ candidates to reject protons.
In criterion (f), we first collect ${e}^{+/-}$ candidate events using the energy deposit sum in the IMC bottom X- and Y-layers \citep{CALope2018}.
We additionally select ${e}^{+/-}$ events using the energy (deposit) concentration.  
The latter is defined as the fraction of energy deposit in $\pm 9$ scintillating fibers surrounding the reconstructed particle track to the total energy deposit in each layer. 
For each candidate ${e}^{+/-}$ and p events in criterion (g), we derive the distribution of the lateral spread parameter $R_E$ defined as,
\begin{equation}
R_E = \sqrt{\dfrac{\Sigma_i \Delta E_i X^2_i}{\Sigma_i \Delta E_i}}
\end{equation}
where $\Delta E_i$ and $X_i$ are the energy deposit and distance from the reconstructed track in the $i$-th lead tungstate crystal in the TASC top layer, respectively. 
The upper panel of Fig.$~$S1 shows examples of the $R_E$ distributions of candidate ${e}^{+/-}$ obtained after criteria (a)-(f) during the Carrington rotation of CR 2208 between September 2 and September 28, 2018, while the lower panel shows that of $p$ obtained after criteria (a)-(e) without (f). 
It is seen that the observed distributions are well reproduced by the MC simulations. 
The reduced chi-square of the difference between the observed and MC $R_E$ distributions shown in Fig. S1 are about 1.14 and 1.04 for ${e}^{+/-}$ and $p$ events, respectively.
We define the events with $R_E$ below 2.0 cm in the upper panel as ${e}^{+/-}$ and the events with $R_E$ above 2.0 cm in the lower panel as $p$. 
This value of $R_E=2.0$ cm is consistent with one Moli\`{e}re unit in the TASC layer. 

\begin{figure}[p]
\centering
\includegraphics[width=25pc]{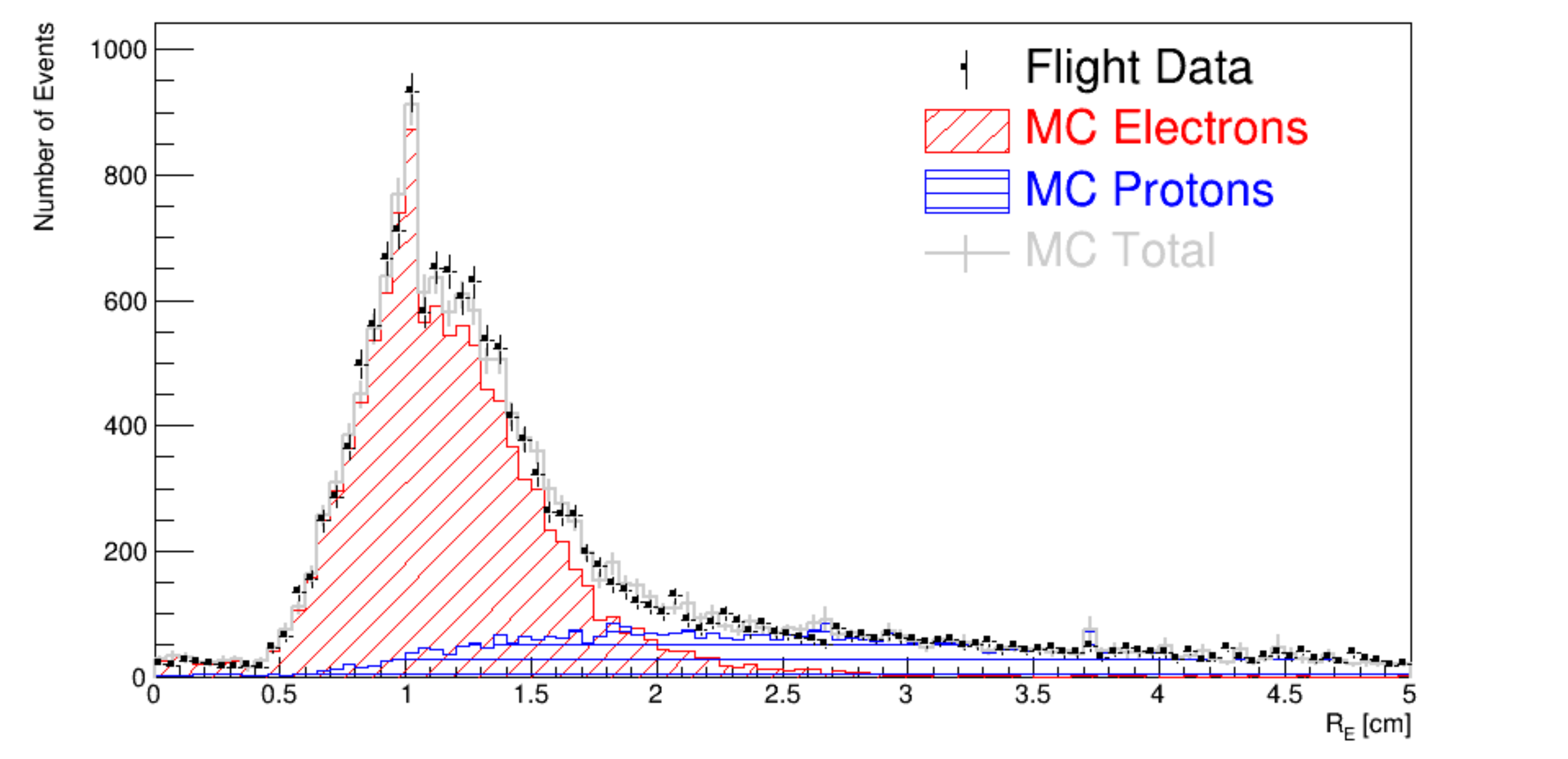}
\includegraphics[width=25pc]{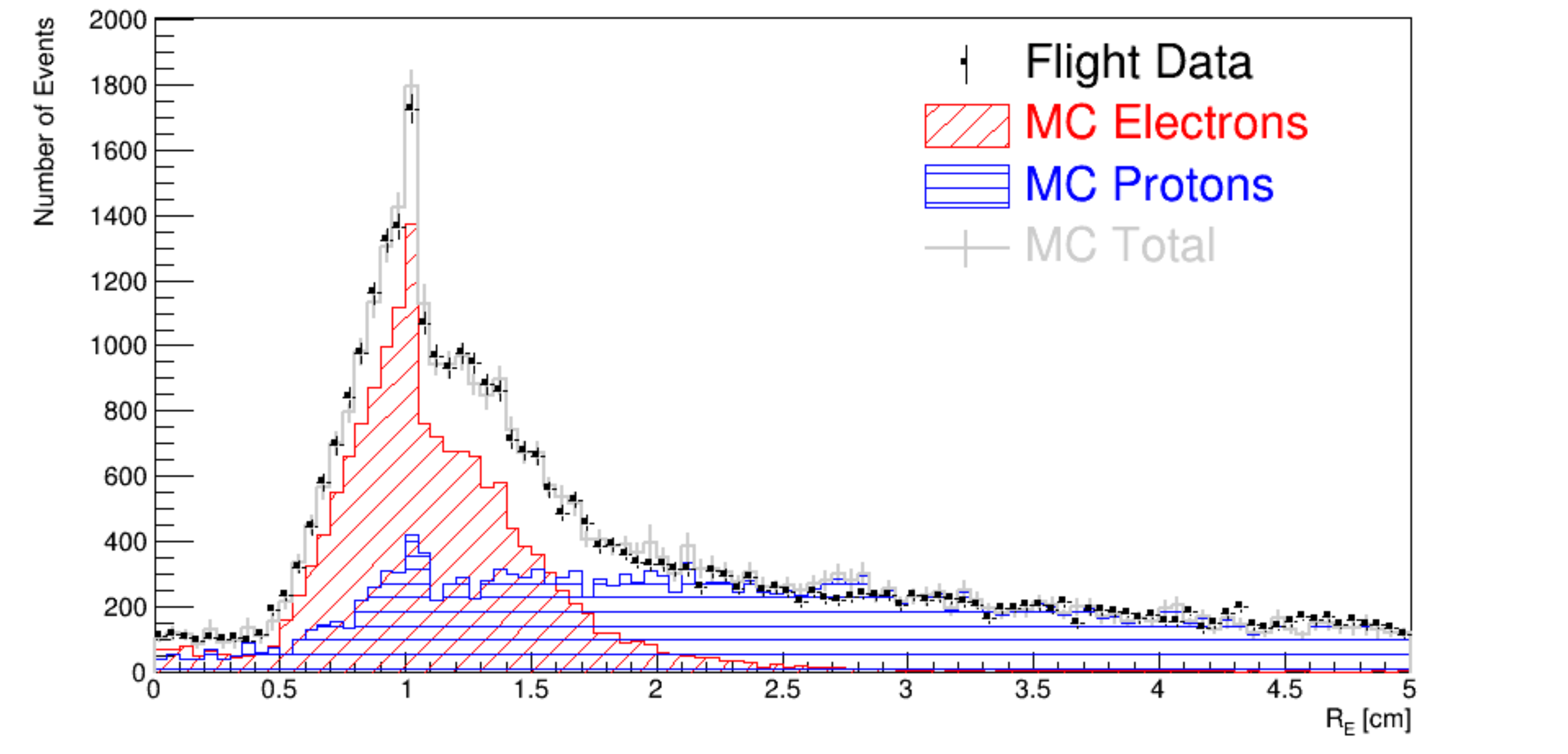}
\caption{\label{Re} $R_E$ distribution of candidate electron (top) and proton (bottom) events during the Carrington rotation of CR 2208.}
\end{figure}

The upper panel of Fig.$~$S2 shows the fractional contamination due to $p$ ($f_{p}$) in the observed ${e}^{+/-}$ count rate ($C_{{e^{+/-}}\_{obs}}$) on the left vertical axis, while the lower panel displays the fractional contamination due to ${e}^{+/-}$ ($f_{e^{+/-}}$) in the observed $p$ count rate ($C_{p\_{obs}}$). 
We can see that $f_{p}$ and $f_{e^{+/-}}$ decrease and increase with decreasing solar activity and reach minimum and maximum values when $C_{{e}^{-}}$, shown in Fig.$~$2(b) in the main text, reaches the maximum in 2020, respectively.
These variations are caused by the drift effect that results in the change in the flux ratio of GCRs with $q<0$ or $q>0$.
We calculate the fractional contamination by positrons ($f_{e^{+}}$) in $C_{{e^{+/-}}\_{obs}}$ by simply assuming that the positron flux is 7 $\%$ of the electron flux in October 2015 as reported by \citet{AMSe2018} and its temporal variation in $\%$ follows the proton flux. 
The upper panel of Fig.$~$S2 also shows the fractional contamination due to positrons calculated in this way on the right vertical axis. 
We calculate $C_{{e^-}}$ and $C_{p}$ using  $f_{e^{+}}$, $f_{p}$, and $f_{e^{+/-}}$ in each Carrington rotation, as 
\begin{equation}
 \begin{array}{lll}
 C_{{e^-}} &=& (1.0-f_{e^{+}})(1.0-f_{p})\times {C_{{e^{+/-}}\_{obs}}}~, \\
 C_{p} &=& (1.0-f_{e^{+/-}})\times {C_{p\_{obs}}}~. \\
 \end{array}
\end{equation}

\begin{figure}[h]
\centering
\includegraphics[width=25pc]{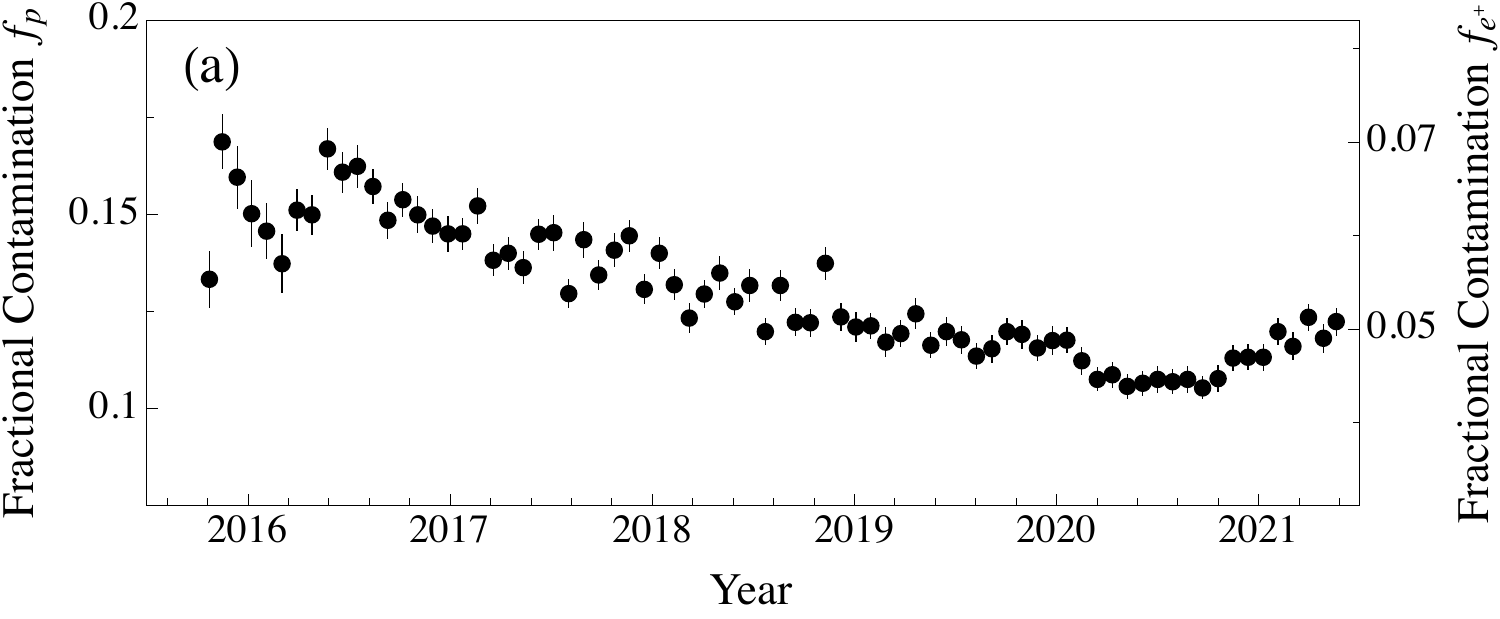}
\includegraphics[width=22.7pc]{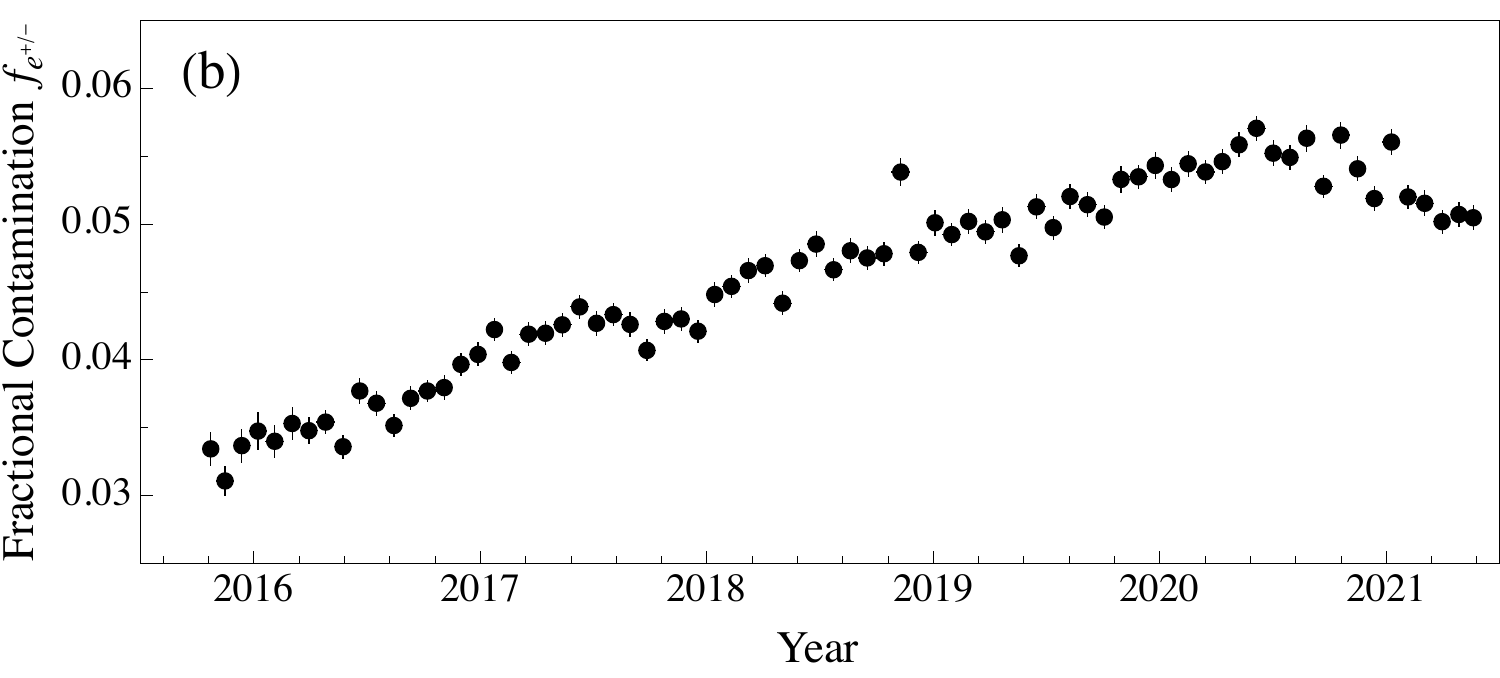}\hspace{1cm}
\caption{(a) Temporal variations of the fractional contamination $f_{p}$ by $p$ in $C_{{e^{+/-}}\_{obs}}$ on the left vertical axis and fractional contamination $f_{e^+}$ by ${e}^+$ in $C_{{e^{+/-}}\_{obs}}$ on the right vertical axis. (b) Electron and positron contamination $f_{e^{+/-}}$ in $C_{p\_{obs}}$ as a function of time.}
\end{figure}

\newpage

\section{Correlation between proton (electron) count rates and neutron monitor count rates}

Figure S3 shows proton (a) and electron (b) count rates at the average rigidity of 3.8 GV each as a function of the Oulu neutron monitor count rate. A good correlation in Fig.$~$S3(a) ensures that our determination of the proton count rate is handled properly. 

\begin{figure}[h]
\centering
\includegraphics[width=22pc]{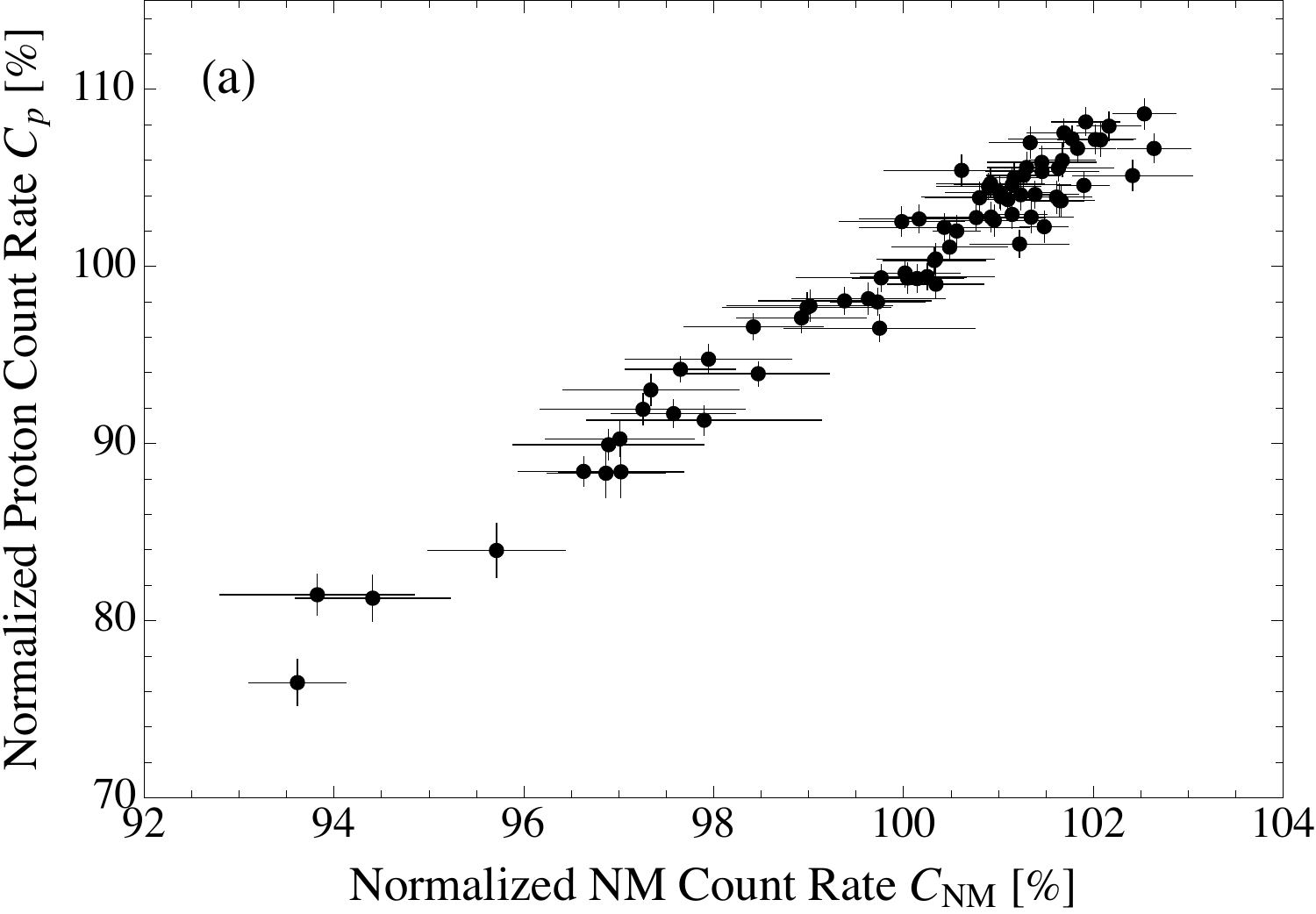}
\vspace{1cm}
\includegraphics[width=22pc]{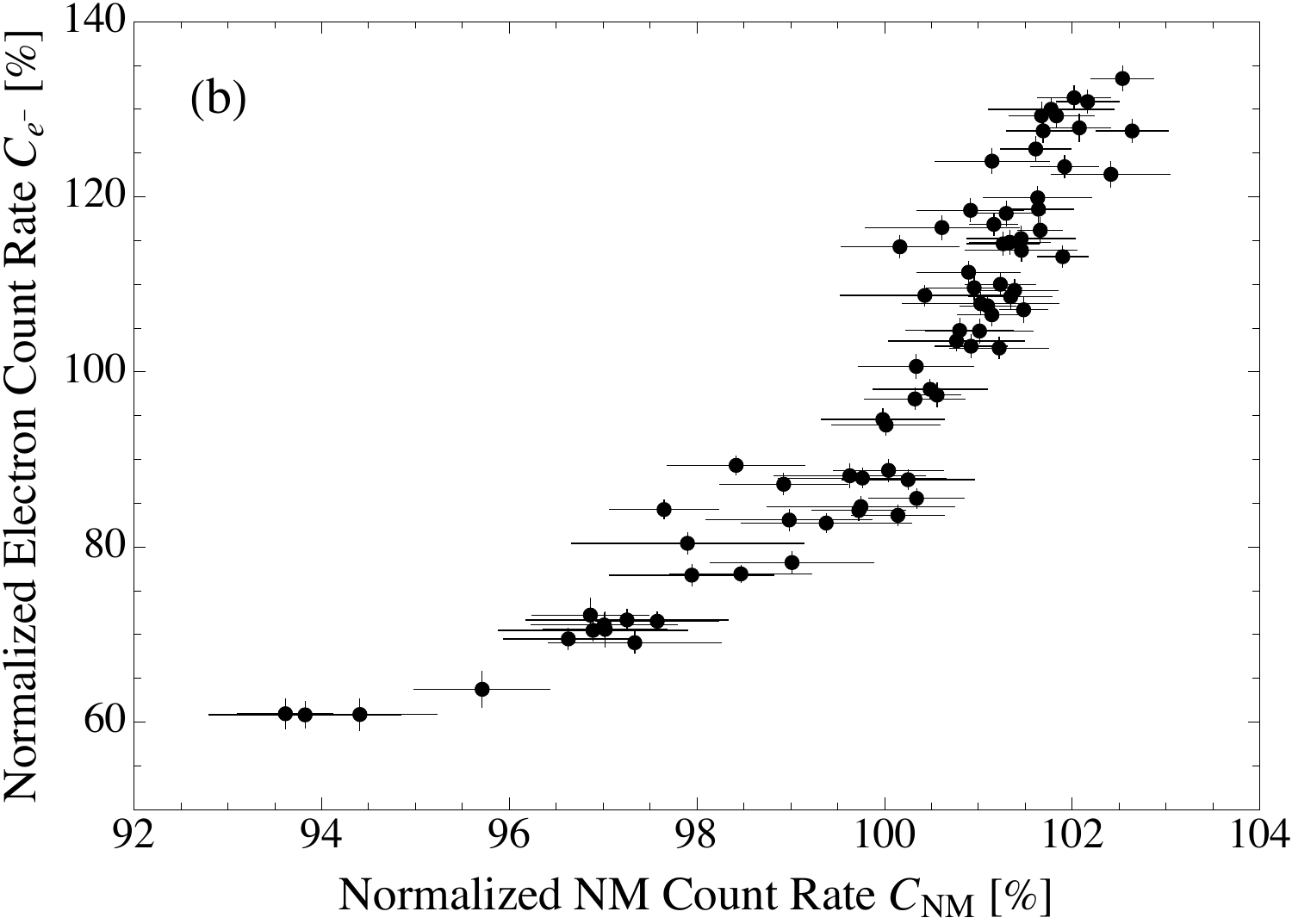}
\caption{CALET proton (a) and electron (b) count rates at the average rigidity of 3.8 GV as a function of neutron monitor count rates $C_{\mathrm{NM}}$ at the Oulu station. Correlation coefficients and regression coefficients of proton (electron) count rates with neutron monitor count rates are 0.98$\pm$0.07 (0.91$\pm$0.07) and 3.32$\pm$0.24 (9.2$\pm$0.7), respectively.}
\end{figure}

\newpage
\section{Normalized Count Rates of electrons and protons}

Table$~$S1 presents the normalized count rates of electrons and protons at the average rigidity of 3.8 GV, fractional contaminations due to protons and positrons in the observed ${e^{+/-}}$, and the fractional contamination due to ${e^{+/-}}$ in the observed protons, for each Carrington rotation.

\begin{table}[h]
\caption{Electron and proton normalized count rates, $C_{{e^-}}$ and $C_{{p}}$ with statistical errors, at the average rigidity of 3.8 GV, and fractional contaminations, $f_{p}$, $f_{e^{+}}$, and $f_{e^{+/-}}$, for each Carrington rotation.     }
\begin{tabular}{C{18mm}C{20mm}C{20mm}C{20mm}C{23mm}C{30mm}c}
\hline
Carrington& Start Date   & $C_{{e^-}}$      & $C_{{p}}$ 
   & \multirow{2}{*}{$f_{p}$} & \multirow{2}{*}{$f_{e^{+}}$} & \multirow{2}{*}{$f_{e^{+/-}}$} \\
Rotation   & \tiny{(month/day/year)} & \tiny{ Value$\pm$Stat. ($\%$)}    & {\tiny Value$\pm$Stat. ($\%$)}  & & & \\
\hline
\hline
 CR 2169 & 10/13/2015 & 60.9$\pm$1.7 & 76.5$\pm$1.3 & 0.133$\pm$0.007 & 0.055$\pm$0.004 & 0.0334$\pm$0.0012 \\
 CR 2170 & 10/31/2015 & 60.8$\pm$1.5 & 81.5$\pm$1.1 & 0.169$\pm$0.007 & 0.070$\pm$0.004 & 0.0311$\pm$0.0011 \\
 CR 2171 & 11/28/2015 & 60.8$\pm$1.8 & 81.3$\pm$1.3 & 0.160$\pm$0.008 & 0.066$\pm$0.004 & 0.0337$\pm$0.0012 \\
 CR 2172 & 12/25/2015 & 63.7$\pm$2.0 & 84.0$\pm$1.5 & 0.150$\pm$0.009 & 0.062$\pm$0.004 & 0.0348$\pm$0.0013 \\
 CR 2173 & 01/21/2016 & 72.2$\pm$1.9 & 88.3$\pm$1.4 & 0.146$\pm$0.007 & 0.060$\pm$0.004 & 0.0340$\pm$0.0012 \\
 CR 2174 & 02/18/2016 & 70.6$\pm$2.0 & 88.4$\pm$1.4 & 0.137$\pm$0.007 & 0.057$\pm$0.004 & 0.0353$\pm$0.0012 \\
 CR 2175 & 03/16/2016 & 71.1$\pm$1.3 & 90.3$\pm$1.0 & 0.151$\pm$0.005 & 0.0627$\pm$0.0034 & 0.0348$\pm$0.0009 \\
 CR 2176 & 04/12/2016 & 69.5$\pm$1.2 & 88.4$\pm$0.8 & 0.150$\pm$0.005 & 0.0622$\pm$0.0033 & 0.0354$\pm$0.0009 \\
 CR 2177 & 05/09/2016 & 69.0$\pm$1.2 & 93.0$\pm$0.9 & 0.167$\pm$0.005 & 0.069$\pm$0.004 & 0.0336$\pm$0.0008 \\
 CR 2178 & 06/06/2016 & 71.6$\pm$1.2 & 91.9$\pm$0.9 & 0.161$\pm$0.005 & 0.0668$\pm$0.0035 & 0.0377$\pm$0.0009 \\
 CR 2179 & 07/03/2016 & 70.5$\pm$1.3 & 89.9$\pm$0.9 & 0.162$\pm$0.005 & 0.067$\pm$0.004 & 0.0368$\pm$0.0009 \\
 CR 2180 & 07/30/2016 & 71.5$\pm$1.1 & 91.7$\pm$0.8 & 0.157$\pm$0.004 & 0.0652$\pm$0.0033 & 0.0352$\pm$0.0008 \\
 CR 2181 & 08/27/2016 & 76.8$\pm$1.2 & 94.8$\pm$0.8 & 0.148$\pm$0.005 & 0.0616$\pm$0.0032 & 0.0372$\pm$0.0008 \\
 CR 2182 & 09/23/2016 & 76.9$\pm$1.0 & 93.9$\pm$0.7 & 0.154$\pm$0.004 & 0.0638$\pm$0.0032 & 0.0377$\pm$0.0008 \\
 CR 2183 & 10/20/2016 & 78.2$\pm$1.2 & 97.8$\pm$0.9 & 0.150$\pm$0.005 & 0.0622$\pm$0.0032 & 0.0380$\pm$0.0008 \\
 CR 2184 & 11/16/2016 & 82.7$\pm$1.1 & 98.0$\pm$0.7 & 0.147$\pm$0.004 & 0.0610$\pm$0.0030 & 0.0397$\pm$0.0008 \\
 CR 2185 & 12/14/2016 & 83.1$\pm$1.2 & 97.7$\pm$0.8 & 0.145$\pm$0.004 & 0.0602$\pm$0.0031 & 0.0404$\pm$0.0008 \\
 CR 2186 & 01/10/2017 & 84.1$\pm$1.1 & 98.0$\pm$0.8 & 0.145$\pm$0.004 & 0.0602$\pm$0.0030 & 0.0422$\pm$0.0008 \\
 CR 2187 & 02/06/2017 & 83.6$\pm$1.2 & 99.3$\pm$0.8 & 0.152$\pm$0.004 & 0.0631$\pm$0.0032 & 0.0398$\pm$0.0008 \\
 CR 2188 & 03/05/2017 & 85.5$\pm$1.1 & 99.0$\pm$0.8 & 0.138$\pm$0.004 & 0.0573$\pm$0.0029 & 0.0419$\pm$0.0008 \\
 CR 2189 & 04/02/2017 & 84.6$\pm$1.1 & 96.5$\pm$0.8 & 0.140$\pm$0.004 & 0.0581$\pm$0.0029 & 0.0420$\pm$0.0008 \\
 CR 2190 & 04/29/2017 & 88.7$\pm$1.3 & 99.3$\pm$0.9 & 0.136$\pm$0.004 & 0.0565$\pm$0.0029 & 0.0426$\pm$0.0008 \\
 CR 2191 & 05/26/2017 & 87.7$\pm$1.1 & 99.4$\pm$0.7 & 0.145$\pm$0.004 & 0.0601$\pm$0.0030 & 0.0439$\pm$0.0008 \\
 CR 2192 & 06/23/2017 & 88.1$\pm$1.3 & 98.2$\pm$0.9 & 0.145$\pm$0.004 & 0.0603$\pm$0.0031 & 0.0427$\pm$0.0009 \\
 CR 2193 & 07/20/2017 & 89.3$\pm$1.1 & 96.6$\pm$0.7 & 0.130$\pm$0.004 & 0.0538$\pm$0.0027 & 0.0433$\pm$0.0008 \\
 CR 2194 & 08/16/2017 & 80.4$\pm$1.2 & 91.3$\pm$0.8 & 0.143$\pm$0.004 & 0.0595$\pm$0.0031 & 0.0426$\pm$0.0009 \\
 CR 2195 & 09/12/2017 & 84.3$\pm$1.0 & 94.2$\pm$0.7 & 0.134$\pm$0.004 & 0.0558$\pm$0.0028 & 0.0407$\pm$0.0008 \\
 CR 2196 & 10/10/2017 & 87.1$\pm$1.2 & 97.1$\pm$0.8 & 0.141$\pm$0.004 & 0.0584$\pm$0.0030 & 0.0428$\pm$0.0009 \\
 CR 2197 & 11/06/2017 & 87.9$\pm$1.2 & 99.3$\pm$0.8 & 0.144$\pm$0.004 & 0.0600$\pm$0.0030 & 0.0430$\pm$0.0008 \\
 CR 2198 & 12/03/2017 & 93.9$\pm$1.2 & 99.6$\pm$0.8 & 0.131$\pm$0.004 & 0.0542$\pm$0.0027 & 0.0421$\pm$0.0008 \\
 CR 2199 & 12/30/2017 & 94.5$\pm$1.2 & 102.5$\pm$0.8 & 0.140$\pm$0.004 & 0.0581$\pm$0.0029 & 0.0448$\pm$0.0009 \\
 CR 2200 & 01/27/2018 & 96.9$\pm$1.2 & 100.3$\pm$0.8 & 0.132$\pm$0.004 & 0.0547$\pm$0.0027 & 0.0454$\pm$0.0008 \\
 CR 2201 & 02/23/2018 & 100.6$\pm$1.3 & 100.4$\pm$0.9 & 0.123$\pm$0.004 & 0.0512$\pm$0.0026 & 0.0466$\pm$0.0009 \\
 CR 2202 & 03/22/2018 & 98.0$\pm$1.1 & 101.1$\pm$0.7 & 0.1294$\pm$0.0035 & 0.0537$\pm$0.0026 & 0.0469$\pm$0.0008 \\
 CR 2203 & 04/19/2018 & 97.3$\pm$1.4 & 102.0$\pm$0.9 & 0.135$\pm$0.004 & 0.0560$\pm$0.0029 & 0.0442$\pm$0.0008 \\
 CR 2204 & 05/16/2018 & 103.5$\pm$1.1 & 102.7$\pm$0.8 & 0.1274$\pm$0.0034 & 0.0529$\pm$0.0026 & 0.0473$\pm$0.0008 \\
 CR 2205 & 06/12/2018 & 104.7$\pm$1.4 & 103.9$\pm$0.9 & 0.132$\pm$0.004 & 0.0546$\pm$0.0028 & 0.0485$\pm$0.0009 \\
 CR 2206 & 07/09/2018 & 107.5$\pm$1.2 & 103.8$\pm$0.8 & 0.1197$\pm$0.0033 & 0.0497$\pm$0.0025 & 0.0466$\pm$0.0008 \\
 CR 2207 & 08/06/2018 & 102.9$\pm$1.3 & 102.8$\pm$0.8 & 0.132$\pm$0.004 & 0.0546$\pm$0.0027 & 0.0480$\pm$0.0009 \\
 CR 2208 & 09/02/2018 & 106.5$\pm$1.2 & 102.9$\pm$0.8 & 0.1221$\pm$0.0035 & 0.0507$\pm$0.0025 & 0.0475$\pm$0.0008 \\
 CR 2209 & 09/29/2018 & 102.7$\pm$1.2 & 101.3$\pm$0.8 & 0.1220$\pm$0.0034 & 0.0507$\pm$0.0025 & 0.0478$\pm$0.0008 \\
 CR 2210 & 10/26/2018 & 104.6$\pm$1.4 & 104.2$\pm$0.9 & 0.137$\pm$0.004 & 0.0570$\pm$0.0029 & 0.0538$\pm$0.0010 \\
 CR 2211 & 11/23/2018 & 107.8$\pm$1.2 & 103.9$\pm$0.8 & 0.1235$\pm$0.0035 & 0.0513$\pm$0.0026 & 0.0479$\pm$0.0008 \\
 CR 2212 & 12/20/2018 & 109.6$\pm$1.4 & 102.6$\pm$0.9 & 0.121$\pm$0.004 & 0.0502$\pm$0.0026 & 0.0501$\pm$0.0009 \\
 CR 2213 & 01/16/2019 & 111.4$\pm$1.2 & 104.5$\pm$0.7 & 0.1212$\pm$0.0032 & 0.0503$\pm$0.0025 & 0.0492$\pm$0.0008 \\
 CR 2214 & 02/13/2019 & 108.6$\pm$1.4 & 102.8$\pm$0.9 & 0.117$\pm$0.004 & 0.0486$\pm$0.0025 & 0.0502$\pm$0.0009 \\
 CR 2215 & 03/12/2019 & 109.3$\pm$1.2 & 104.1$\pm$0.8 & 0.1192$\pm$0.0033 & 0.0495$\pm$0.0024 & 0.0494$\pm$0.0008 \\
 CR 2216 & 04/08/2019 & 107.1$\pm$1.4 & 102.2$\pm$0.9 & 0.124$\pm$0.004 & 0.0516$\pm$0.0027 & 0.0503$\pm$0.0009 \\
 CR 2217 & 05/05/2019 & 108.7$\pm$1.2 & 102.2$\pm$0.8 & 0.1162$\pm$0.0032 & 0.0482$\pm$0.0024 & 0.0477$\pm$0.0008 \\
\hline
\end{tabular}
\\ \leftline{\hspace{1cm} \it{Continued on next page}}
\end{table}
 
\setcounter{table}{0}

\newpage

\begin{table}[h]
\caption{--- \it{Continued from previous page} }
\begin{tabular}{C{18mm}C{20mm}C{20mm}C{20mm}C{23mm}C{30mm}c}
\hline
Carrington& Start Date   & $C_{{e^-}}$      & $C_{{p}}$ 
   & \multirow{2}{*}{$f_{p}$} & \multirow{2}{*}{$f_{e^{+}}$} & \multirow{2}{*}{$f_{e^{+/-}}$} \\
Rotation   & \tiny{(month/day/year)} & \tiny{ Value$\pm$Stat. ($\%$)}    & {\tiny Value$\pm$Stat. ($\%$)}  & & & \\
\hline
\hline
 CR 2218 & 06/02/2019 & 110.0$\pm$1.3 & 104.0$\pm$0.8 & 0.1197$\pm$0.0035 & 0.0497$\pm$0.0025 & 0.0513$\pm$0.0009 \\
 CR 2219 & 06/29/2019 & 114.6$\pm$1.3 & 105.1$\pm$0.8 & 0.1176$\pm$0.0034 & 0.0488$\pm$0.0025 & 0.0497$\pm$0.0008 \\
 CR 2220 & 07/26/2019 & 116.8$\pm$1.2 & 105.0$\pm$0.8 & 0.1134$\pm$0.0031 & 0.0471$\pm$0.0023 & 0.0520$\pm$0.0008 \\
 CR 2221 & 08/23/2019 & 118.1$\pm$1.4 & 105.6$\pm$0.9 & 0.1153$\pm$0.0035 & 0.0479$\pm$0.0024 & 0.0514$\pm$0.0009 \\
 CR 2222 & 09/19/2019 & 113.1$\pm$1.2 & 104.6$\pm$0.8 & 0.1197$\pm$0.0033 & 0.0497$\pm$0.0025 & 0.0505$\pm$0.0008 \\
 CR 2223 & 10/16/2019 & 116.2$\pm$1.4 & 103.7$\pm$0.9 & 0.119$\pm$0.004 & 0.0494$\pm$0.0025 & 0.0533$\pm$0.0009 \\
 CR 2224 & 11/12/2019 & 119.9$\pm$1.3 & 105.5$\pm$0.8 & 0.1155$\pm$0.0032 & 0.0480$\pm$0.0024 & 0.0535$\pm$0.0008 \\
 CR 2225 & 12/10/2019 & 118.6$\pm$1.5 & 103.7$\pm$0.9 & 0.117$\pm$0.004 & 0.0487$\pm$0.0025 & 0.0543$\pm$0.0009 \\
 CR 2226 & 01/06/2020 & 123.4$\pm$1.3 & 108.2$\pm$0.8 & 0.1175$\pm$0.0031 & 0.0488$\pm$0.0024 & 0.0533$\pm$0.0008 \\
 CR 2227 & 02/02/2020 & 122.5$\pm$1.4 & 105.1$\pm$0.8 & 0.1122$\pm$0.0034 & 0.0466$\pm$0.0024 & 0.0544$\pm$0.0009 \\
 CR 2228 & 02/29/2020 & 130.9$\pm$1.3 & 107.9$\pm$0.8 & 0.1074$\pm$0.0029 & 0.0446$\pm$0.0022 & 0.0538$\pm$0.0009 \\
 CR 2229 & 03/28/2020 & 127.5$\pm$1.4 & 106.7$\pm$0.8 & 0.1086$\pm$0.0031 & 0.0451$\pm$0.0023 & 0.0546$\pm$0.0009 \\
 CR 2230 & 04/24/2020 & 133.5$\pm$1.4 & 108.6$\pm$0.9 & 0.1056$\pm$0.0031 & 0.0438$\pm$0.0022 & 0.0558$\pm$0.0009 \\
 CR 2231 & 05/21/2020 & 131.3$\pm$1.3 & 107.2$\pm$0.8 & 0.1064$\pm$0.0030 & 0.0442$\pm$0.0022 & 0.0570$\pm$0.0009 \\
 CR 2232 & 06/18/2020 & 127.9$\pm$1.6 & 107.1$\pm$0.9 & 0.1074$\pm$0.0034 & 0.0446$\pm$0.0023 & 0.0552$\pm$0.0009 \\
 CR 2233 & 07/15/2020 & 129.2$\pm$1.3 & 106.7$\pm$0.8 & 0.1068$\pm$0.0030 & 0.0443$\pm$0.0022 & 0.0549$\pm$0.0009 \\
 CR 2234 & 08/11/2020 & 129.2$\pm$1.5 & 106.0$\pm$0.9 & 0.1074$\pm$0.0034 & 0.0446$\pm$0.0023 & 0.0563$\pm$0.0010 \\
 CR 2235 & 09/07/2020 & 130.0$\pm$1.2 & 107.2$\pm$0.7 & 0.1052$\pm$0.0028 & 0.0437$\pm$0.0021 & 0.0528$\pm$0.0008 \\
 CR 2236 & 10/05/2020 & 125.4$\pm$1.4 & 103.9$\pm$0.9 & 0.1076$\pm$0.0033 & 0.0447$\pm$0.0023 & 0.0565$\pm$0.0010 \\
 CR 2237 & 11/01/2020 & 127.5$\pm$1.3 & 107.5$\pm$0.8 & 0.1129$\pm$0.0031 & 0.0469$\pm$0.0023 & 0.0540$\pm$0.0009 \\
 CR 2238 & 11/28/2020 & 118.4$\pm$1.3 & 104.7$\pm$0.8 & 0.1131$\pm$0.0032 & 0.0470$\pm$0.0024 & 0.0519$\pm$0.0009 \\
 CR 2239 & 12/25/2020 & 124.1$\pm$1.4 & 104.6$\pm$0.9 & 0.1131$\pm$0.0033 & 0.0470$\pm$0.0024 & 0.0560$\pm$0.0009 \\
 CR 2240 & 01/22/2021 & 114.3$\pm$1.3 & 102.7$\pm$0.8 & 0.1197$\pm$0.0034 & 0.0497$\pm$0.0025 & 0.0520$\pm$0.0008 \\
 CR 2241 & 02/18/2021 & 116.5$\pm$1.4 & 105.4$\pm$0.9 & 0.1159$\pm$0.0034 & 0.0481$\pm$0.0024 & 0.0515$\pm$0.0009 \\
 CR 2242 & 03/17/2021 & 113.9$\pm$1.2 & 105.3$\pm$0.8 & 0.1234$\pm$0.0033 & 0.0512$\pm$0.0025 & 0.0502$\pm$0.0008 \\
 CR 2243 & 04/14/2021 & 115.2$\pm$1.4 & 105.9$\pm$0.9 & 0.118$\pm$0.004 & 0.0490$\pm$0.0025 & 0.0507$\pm$0.0009 \\
 CR 2244 & 05/11/2021 & 114.8$\pm$1.3 & 107.0$\pm$0.9 & 0.1223$\pm$0.0035 & 0.0508$\pm$0.0025 & 0.0505$\pm$0.0009 \\
\hline
\end{tabular}
\end{table}

\newpage

\section{\label{sec:calc}Numerical Drift Model of the Solar Modulation}

The transport of GCRs in the heliosphere is described by the Fokker-Planck equation,
\begin{equation}
\frac{\partial f}{\partial t}=\bm{\nabla}\cdot(\bm{\kappa}\cdot\bm{\nabla}f-\bm{V}_{\mathrm{SW}}f)+\frac{1}{3p^{2}}(\bm{\nabla}\cdot\bm{V}_{\mathrm{SW}})\frac{\partial}{\partial p}(p^{3}f),
\label{eq:cd}
\end{equation}
where $f(\bm{r},p,t)$ is the phase space distribution function of GCRs with momentum $p$ at position $\bm{r}$ and time $t$,  $\bm{V}_{\mathrm{SW}}$ is the solar wind speed and $\bm{\kappa}$ is the spatial diffusion tensor. We obtain the solution $f$ at Earth by calculating the following set of the stochastic differential equations (SDEs) which are mathematically equivalent to Eq.$~$(S3) \citep{Yamada1998,Zhang1999,Pei2010,Strauss2012,Miyake2017},
\begin{equation}
\begin{array}{llllll}
d{\bm{r}} & = & ( \bm{\nabla} \cdot {\bm{\kappa}}
  + {\bm{V}_{\mathrm{SW}}} + {\bm{V}_{\mathrm{D}}} )dt
  + \sum_{s} {\bm{\sigma}}_{s} dW_{s}(t)~, \\
   & &\\
dp & = &-\dfrac{1}{3} p(\bm{\nabla} \cdot {\bm{V}_{\mathrm{SW}}})dt~,
\end{array}\label{sde}
\end{equation}
where $\bm{V}_{\mathrm{D}}$ is the gradient-curvature drift velocity, ${\bm{\sigma}}_{s}$ is the tensor which gives $\sum_{s} {\sigma}^{\mu}_{s} {\sigma}^{\nu}_{s} = 2 {\kappa}^{\mu \nu}$ with ${\kappa}^{\mu \nu}$ denoting the ${\mu \nu}$ element of $\bm{\kappa}$, $dW_{s}$ is a vector denoting the random-walk along the $s$-axis according to the Gaussian probability distribution of the Wiener process \citep{Gardiner1989}.\\

The diffusion tensor $\bm{\kappa}$ in a coordinate system in which the x-axis points away from the Sun along the magnetic field line, the y-axis is directed southward perpendicular to a plane including the magnetic field line and the radial vector, and the z-axis completes the right-handed coordinated system, is written as
\begin{equation}
\boldsymbol{\kappa} = \left(
\begin{array}{ccc}
\kappa_{\parallel} & 0 & 0 \\
0 & \kappa_{\bot 1} &0 \\
0 & 0 & \kappa_{\bot 2}
\end{array}
\right)~.
\end{equation}
We define the diffusion coefficients in terms of the particle's velocity ($v$) and momentum ($p$) and the strength of the heliospheric magnetic field (HMF) ($B$) at the particle's location, as
\begin{equation}
 \begin{array}{llllll}
 \kappa_{\parallel} &=& \kappa_{\parallel}^0 \beta 
 	\left( \dfrac{p}{1 \mathrm{GeV/c}} \right)^{a}
 	\left( \dfrac{B_{\mathrm{E}}}{B} \right)^{b_{\parallel}}~,\\
 \kappa_{\bot 1} &=& \kappa_{\bot 1}^0 \beta 
 	\left( \dfrac{p}{1 \mathrm{GeV/c}} \right)^{a}
 	\left( \dfrac{B_{\mathrm{E}}}{B} \right)^{b_{\bot 1}}~,\\
 \kappa_{\bot 2} &=& \kappa_{\bot 2}^0 \beta
 	\left( \dfrac{p}{1 \mathrm{GeV/c}} \right)^{a}
 	\left( \dfrac{B_{\mathrm{E}}}{B} \right)^{b_{\bot 2}}~,
  \end{array}
\label{kappa}
\end{equation}
where $\beta$ is the particle's velocity relative to the speed of light ($c$) and $B_{\mathrm{E}}$ is $B$ at Earth. $\kappa_{\parallel}^0$, $\kappa_{\bot 1}^0$, $\kappa_{\bot 2}^0$, $a$, $b_{\parallel}$, $b_{\bot 1}$, and $b_{\bot 2}$ are free parameters which are set to $\kappa_{\parallel}^0 = 1\times 10^{22} ~\mathrm{cm^2/s}$, $\kappa_{\bot 1}^0= 2\times 10^{20} ~ \mathrm{cm^2/s}$, $\kappa_{\bot 2}^0= 1\times 10^{20} ~ \mathrm{cm^2/s}$, $a=1.0$, $b_{\parallel}=1.0$, $b_{\bot 1}=1.4 \mathrm{cos}^2\theta+0.6 \mathrm{sin}^2\theta$, and $b_{\bot 2}=1.0$, respectively, where $\theta$ is the colatitude of the particle's location in the heliocentric polar coordinate system, for the best reproduction of the GCR energy spectra observed at Earth \citep{MiyakeICRC2017}.\\

The $\bm{V}_{\mathrm{D}}$ of a particle in the HMF $\bm{B}$ is given as \citep{IJ1979}
\begin{equation}
\bm{V}_{\mathrm{D}}=\dfrac{pvc}{3qB^4} \left[ 
B^2(\bm{\nabla} \times \bm{B}) + \bm{B} \times \bm{\nabla}B^2 \right],
\label{driftV}
\end{equation}
where $q$ is the particle's charge. To calculate $\bm{V}_{\mathrm{D}}$ in Eq.$~$(S7) at the particle's location, we adopt the Parker-Spiral HMF in the heliocentric polar coordinate system formed with unit vectors ($\bm{e}_r$, $\bm{e}_\theta$, $\bm{e}_\phi$) defined as
\begin{equation}
\bm{B}=B_{\mathrm{E}}(\dfrac{r_{\mathrm{E}}}{r})^2\left( \bm{e}_r-\dfrac{r \Omega_{\odot} \sin \theta}{V_{\mathrm{SW}}}\bm{e}_\phi \right)\left[1-2H(\theta-\theta') \right],
\label{ParkerB}
\end{equation}
where $r$ is the radial distance from the Sun, $B_{\mathrm{E}}$ is the strength of $\bm{B}$ observed at Earth where $r=r_{\mathrm{E}}$ and $\theta = 90^\circ$, $\Omega_{\odot}$ is the angular velocity of the solar rotation and $H$ is the Heaviside step function denoting the abrupt reversal of the $\bm{B}$ orientation across the Heliospheric Current Sheet (HCS) at $\theta=\theta'$.  In the present model, $\bm{V}_{\mathrm{SW}}$ is assumed to be radial and constant during each solar rotation period (with its magnitude calculated from the rotation average of $V_{\mathrm{SW}}$ observed at Earth). We model the HCS location ($\theta'$) at $r$, $\phi$ and $t$, as
\begin{equation}
\theta'=\dfrac{\pi}{2}-\sin^{-1} \left[ \sin \alpha \sin \left( \phi-\phi_E-\Omega_{\odot}t+\dfrac{(r-r_{\odot})\Omega_{\odot} }{V_{\mathrm{SW}}} \right) \right],
\end{equation}
where $\phi$ is the heliospheric azimuth angle, $\phi_E$ is $\phi$ of Earth, $r_{\odot}$ is the radius of the Sun, and $\alpha$ is the HCS tilt angle which is assumed to be constant during each solar rotation period. Taking account of the solar wind propagation to the particle's location, we calculate $\bm{B}$ at $\bm{r}$ and $t$ from $B_{\mathrm{E}}$ observed at Earth at $t-r/V_{\mathrm{SW}}$ by Eq.$~$(S9). It is noted that Eq.$~$(S7) cannot be used to calculate the $\bm{V}_{\mathrm{D}}$ of particles crossing the HCS where $\bm{B}$ reverses discontinuously. The modeling of the drift velocity of particles meandering along the HCS is very crucial, because it is expected to be much faster than the solar wind and the typical amount of diffusion \citep{KJ1983,Burger1989}. 
By including a modification to the model by \citet{Burger1989} that makes it insensitive to the structure of the HCS as the particle's Larmor radius ($R_{\mathrm{L}}$) increases, we calculate the magnitude of the drift velocity along the HCS ($V_{\mathrm{HCS}}$), as
\begin{equation}
V_{\mathrm{HCS}} = \left[0.457-0.412\left(\dfrac{d}{\lambda_{\mathrm{HCS}} R_{\mathrm{L}}}\right)+0.0915\left(\dfrac{d}{\lambda_{\mathrm{HCS}} R_{\mathrm{L}}}\right)^2 \right]
\lambda_{\mathrm{HCS}} v~,
\end{equation}
with
\begin{equation}
\lambda_{\mathrm{HCS}} = \left\{
\begin{array}{cccc}
&1~&,~~& \mathrm{if}~ R_{\mathrm{L}} \leq V_{\mathrm{SW}}T_{\mathrm{S}} \\
&&\\
&\dfrac{V_{\mathrm{SW}}T_{\mathrm{S}}}{R_{\mathrm{L}}}~&,
~~& \mathrm{if}~ R_{\mathrm{L}} > V_{\mathrm{SW}}T_{\mathrm{S}}
\end{array}
\right.~,
\end{equation}
where $d$ is the minimum distance between the particle and the HCS, and $T_{\mathrm{S}}$ is the rotation period of the Sun. The above $V_{\mathrm{HCS}}$ is applied only for particles near the HCS within $d/R_{\mathrm{L}}<2\lambda_{\mathrm{HCS}}$.

The second equation of Eq.$~$(S4) denotes the adiabatic cooling of GCRs in an expanding solar wind. By reversing the sign of the velocity ($\bm{V}_{\mathrm{SW}}$ and $\bm{V}_{\mathrm{D}}$), we solve Eq.$~$(S4) backward in time and trace the particle's trajectory in phase space from Earth at $\bm{r}_{\mathrm{E}}$ until the particle arrives back at the outer boundary at $\bm{r}_{\mathrm{out}}$, which is assumed to be a sphere centered at the Sun with a radius of 100 AU. 
By repeating this calculation for a sufficient number of particles, we obtain a distribution function $F(\bm{r}_{\mathrm{E}}, p_{\mathrm{E}} | \bm{r}_{\mathrm{out}}, p)$ which gives the probability of a particle with $p_{\mathrm{E}}$ at Earth to reach the outer boundary with $p$. We then calculate the phase space distribution of GCRs $f_{\mathrm{E}} (p_{\mathrm{E}})$ at Earth, as
\begin{equation}
f_{E}(p_{\mathrm{E}}) = \int f_{\mathrm{LIS}}(p)F( p_{\mathrm{E}}, \bm{r}_{\mathrm{E}}~|~p, \bm{r}_{\mathrm{out}}) dp~,
\label{distf}
\end{equation}
where $f_{\mathrm{LIS}}(p)$ is the phase space distribution of GCRs in local interstellar space. 
We calculate $f_{\mathrm{LIS}}(p)$ by using the following empirical LIS spectra of GCRs, $J_{p}(E)=f_{\mathrm{LIS}_{p}}(p)p^2$ for GCR protons and $J_{e^-}(E)=f_{\mathrm{LIS}_{e^-}}(p)p^2$ for GCR electrons.
We adopt $J_{p}(E)$ as a function of the kinetic energy $E$, 
\begin{equation}
\begin{array}{llllll}
J_{p}(E) = 16.0 \left( 1+\dfrac{4.2}{E^{1.22}}+\dfrac{1.3}{E^{2.8}}+\dfrac{0.0087}{E^{4.32}} \right) ^{-1} E^{-2.73}~.\end{array}\label{lis}
\end{equation}
We also adopt the $J_{e^-}(E)$ proposed by Potgieter \cite{Potgieter2015} with an additional modulation of $\phi=350~\mathrm{MV}$ based on the force-field approximation \cite{Gleeson1968}. 
The additional force-field modulation is due to the modelling of solar modulation in the heliosheath.

\bibliography{caletSMdraft_ver10}